\documentclass[12pt,a4paper]{article}
\usepackage{amsmath} 
\usepackage{amssymb}
\usepackage{graphicx,epsfig}
\usepackage[mathscr]{eucal}
\usepackage{color} 
\usepackage[numbers,sort&compress]{natbib}
\usepackage{setspace}
\usepackage{xspace}

\usepackage{multirow}

\def\ltap{\raisebox{-.4ex}{\rlap{$\,\sim\,$}} \raisebox{.4ex}{$\,<\,$}} 
\def\gtap{\raisebox{-.4ex}{\rlap{$\,\sim\,$}} \raisebox{.4ex}{$\,>\,$}}

\newcommand\as{\alpha_{\mathrm{S}}} 
\newcommand\f[2]{\frac{#1}{#2}} 
 
\def\to{\rightarrow} 
\def\nn{\nonumber}

\def\ms{\ensuremath{{\overline {\rm MS}}}\xspace}

\def\Mt{\ensuremath{{M_{t}}}\xspace}
\def\mtt{\ensuremath{m_{t{\bar t}}}\xspace}

\def\ttb{\ensuremath{t {\bar t}}\xspace}

\def\mbar{\overline{m}_t}
\def\mbarlam{\overline{m}_\lambda}

\def\mmunew{m_t(\mum)}
\def\lmnew{{L_{\mmunew} } }
\def\muR{\ensuremath{\mu^{}_R}\xspace}
\def\muF{\ensuremath{\mu^{}_F}\xspace}
\def\mum{\ensuremath{\mu^{}_m}\xspace}
\def\muRsq{\ensuremath{\mu^{2}_R}\xspace}

\def\mumsq{\ensuremath{\mu^{2}_m}\xspace}
\def\z#1{\zeta_{#1}}

\newcommand\Matrix{{\sc Matrix}\xspace}
\newcommand\Munich{{\sc Munich}\xspace}
\newcommand\OpenLoops{{\sc OpenLoops}\xspace}
\newcommand\CRunDec{{\sc CRunDec}\xspace}
\newcommand\Hathor{{\sc Hathor}\xspace}

\newcommand{\reffig}[1]{Fig.~\ref{#1}}

\usepackage{array}
\newcolumntype{L}[1]{>{\raggedright\let\newline\\\arraybackslash\hspace{0pt}}m{#1}}
\newcolumntype{C}[1]{>{\centering\let\newline\\\arraybackslash\hspace{0pt}}m{#1}}
\newcolumntype{R}[1]{>{\raggedleft\let\newline\\\arraybackslash\hspace{0pt}}m{#1}}

\usepackage[hidelinks]{hyperref}
\usepackage{footnotebackref}

\begin{document} 
\begin{titlepage}
\begin{flushright}
ZU-TH 10/20\\
MPP-2020-52
\end{flushright}

\renewcommand{\thefootnote}{\fnsymbol{footnote}}
\vspace*{0.2cm}

\begin{center}
  {\Large \bf Top-quark pair hadroproduction at NNLO:\\[0.2cm] differential predictions with the $\overline{\text{MS}}$ mass}
\end{center}

\par \vspace{2mm}
\begin{center}
  {\bf Stefano Catani${}^{(a)}$, Simone Devoto${}^{(b)}$, Massimiliano Grazzini${}^{(b)}$,\\[0.2cm]
    Stefan Kallweit${}^{(c)}$}
and    
{\bf Javier Mazzitelli${}^{(d)}$}

\vspace{5mm}

${}^{(a)}$INFN, Sezione di Firenze and
Dipartimento di Fisica e Astronomia,\\[0.1cm] 
Universit\`a di Firenze,
I-50019 Sesto Fiorentino, Florence, Italy\\[0.25cm]

${}^{(b)}$Physik Institut, Universit\"at Z\"urich, CH-8057 Z\"urich, Switzerland\\[0.25cm]

$^{(c)}$Dipartimento di Fisica, Universit\`{a} degli Studi di Milano-Bicocca and\\[0.1cm] INFN, Sezione di Milano-Bicocca,
I-20126, Milan, Italy\\[0.25cm]

${}^{(d)}$Max-Planck Institut f\"ur Physik, F\"ohringer Ring 6, 80805 M\"unchen, Germany

\vspace{5mm}

\end{center}

\par \vspace{1mm}
\begin{center} {\large \bf Abstract} 

\end{center}
\begin{quote}
\pretolerance 10000

We consider top-quark pair production at the LHC within the \ms scheme for the renormalisation of the top-quark mass, and we present predictions for total and differential cross sections at next-to-next-to-leading order (NNLO) in QCD. Our state-of-the-art calculation extends the available differential results by one order in the perturbative expansion, and it is relevant for a precise determination of the top-quark mass, including possible effects of the running mass in the \ms scheme. We consider variations of the scale at which the \ms mass of the top quark is evaluated, extending the usual 7-point to a 15-point scale variation. This additional variation is crucial for a reliable estimate of the theoretical uncertainties, especially at low perturbative orders and close to the production threshold of the top-quark pair. We also compute, for the first time, the invariant-mass distribution of the top-quark pair by using a running mass, evaluated at a dynamic scale. Our predictions for the invariant-mass distribution in the \ms scheme are compared with a recent measurement performed by the CMS Collaboration. We observe that the inclusion of the NNLO corrections improves the agreement with the data, and we discuss effects due to the QCD running of the \ms mass of the top quark.

\end{quote}

\vspace*{\fill}
\begin{flushleft}
May 2020
\end{flushleft}
\end{titlepage}

\section{Introduction}
\label{sec:intro}

The top-quark mass is a fundamental parameter of the Standard Model (SM). 
Its large size, of the order of the electroweak (EW) scale, is associated with a strong coupling to the Higgs boson, and therefore with a possible role of the top quark in EW symmetry breaking.
Its precise value directly affects the stability of the EW vacuum in the SM, since top-quark contributions drive the evolution of the Higgs boson self-coupling. In addition, the top-quark mass crucially enters in the computation of radiative corrections to precision SM observables. 

For the above reasons, a precise measurement of the top-quark mass is of utmost importance for the LHC and future collider facilities.
The measurement and interpretation of the top-quark mass at hadron colliders is, however, quite controversial (for related reviews see, e.g., Refs.~\cite{Azzi:2019yne,Hoang:2020iah}).
The top-quark mass is a parameter of the underlying field theory, and this implies that, at the formal level, it has to be treated similarly to any other bare parameter of the SM.
The top-quark mass has to be renormalised and, in particular, its meaning and value depend upon the adopted renormalisation scheme.

A widely used renormalisation scheme for the top-quark mass is the \textit{pole} scheme.
Within this scheme the renormalisation procedure fixes the pole of the quark propagator, at any order in perturbation theory, to the same value, which is the pole mass \Mt.
Other mass renormalisation procedures, such as the \ms scheme (which is the customary scheme for the renormalisation of the QCD coupling $\as$), can be used.
In the \ms scheme, the ultraviolet divergences are renormalised by removing only the singular contributions in the dimensionally regularized formulation of the underlying field theory. In this scheme the pole of the quark propagator receives corrections at any order in perturbation theory and, therefore, the \ms renormalised mass $m_t(\mum)$ differs from the pole mass \Mt. Moreover, the \ms mass depends on the auxiliary renormalisation scale \mum, while the pole mass is renormalisation-scale independent.
The physical predictions of the theory are independent of the scheme, provided the two renormalised masses and the corresponding perturbative calculations are formally related to all orders in perturbation theory. However, the use of different mass renormalisation schemes can have a non-negligible quantitative impact in the context of fixed-order predictions, especially at low perturbative orders.

In the context of top-quark production at hadron colliders, the top 
quark is viewed as a `physical', though unstable, particle.
In the limit of vanishing width of the top quark, this picture 
directly leads to considering theoretical calculations for the production
of \textit{on-shell} top quarks ($t$) and antiquarks (${\bar t}$) with a 
definite pole mass.
The main source of top-quark events at hadron colliders is the 
production of top-quark pairs.
QCD radiative corrections to the \ttb total cross section are 
available up to next-to-next-to-leading order (NNLO) within the pole 
mass scheme~\cite{Baernreuther:2012ws, Czakon:2012zr, Czakon:2012pz, 
Czakon:2013goa, Catani:2019iny}.
In Refs.~\cite{Czakon:2015owf,Czakon:2016ckf,Czakon:2017dip,Catani:2019hip} the NNLO QCD calculation of \ttb production is extended to the computation of differential cross sections.

Considering the calculation of on-shell \ttb production, the pole 
mass of the top quark can be re-expressed at the formal level
in terms of a different mass parameter, such as the \ms mass of the top quark.
Approximate NNLO results for the \ttb total cross section using 
the \ms mass were presented in Refs.~\cite{Langenfeld:2009wd,Ahrens:2011px}.
Using the results of Refs.~\cite{Baernreuther:2012ws,Czakon:2012zr, 
Czakon:2012pz,Czakon:2013goa}, the calculation was later extended
to complete NNLO, and first next-to-leading order (NLO) results for 
differential distributions were presented in Ref.~\cite{Dowling:2013baa}.
A common aspect of the studies of Refs.~\cite{Langenfeld:2009wd,Ahrens:2011px,Dowling:2013baa} and of related calculations (e.g., those used in Refs.~\cite{Sirunyan:2018goh,Sirunyan:2019jyn})
is that, in general, only a fixed 
renormalisation scale is used
for the \ms mass. Specifically, the renormalisation scale is set to 
the value $\mum=\mbar$, where the mass parameter $\mbar$
is defined by the relation $m_t(\mbar)=\mbar$ in terms of the \ms 
mass.
Eventually, in these calculations
the pole mass \Mt is perturbatively re-expressed in terms of the 
scale-independent parameter $\mbar$ in the \ms scheme.
In the journal version of Ref.~\cite{Dowling:2013baa} (see Figs.~3 and 8 therein) effects of $\mu_m$ variations around $\mu_m=\mbar$ are considered only for the cases of the total cross section and the transverse-momentum distribution of the top quark.

In this paper we present, for the first time, a fully differential QCD 
calculation of the top-quark pair cross section at NNLO by using the 
\ms mass.
The results are based on our implementation of the \ttb production 
cross section presented in Refs.~\cite{Catani:2019iny,Catani:2019hip}, 
where fully differential predictions were obtained in the pole scheme 
by using the $q_T$-subtraction method~\cite{Catani:2007vq}.
In addition, and at variance with previous works, we do not consider 
only one fixed renormalisation scale for the evaluation of the \ms mass.
We extensively study \mum variations and their effect on the estimate 
of theoretical uncertainties.
The use of the \ms scheme or, more generally, of a short-distance and 
scale-dependent renormalisation procedure of the top-quark mass
can have potential theoretical advantages with respect to the use of a 
fixed (scale-independent) top-quark mass. Indeed, the scale-dependent 
mass can be evaluated
at the physical scale that is relevant for the observable under consideration.
Therefore, in this paper we also study the use of dynamic scales for 
the running \ms mass.
Specifically, we compute the invariant-mass distribution of the \ttb 
pair by using fixed and dynamic scales, and we compare our NNLO results
with a recent measurement performed by the CMS Collaboration~\cite{Sirunyan:2019jyn}.

The paper is organized as follows.
In Section~\ref{sec:msbar} we introduce the \ms scheme for the renormalisation of the top-quark mass, we discuss its relation with the pole scheme, and we present the relevant formulas for the calculation of the NNLO cross section.
In Section~\ref{sec:resu} we present our numerical results for $t {\bar t}$ production at the LHC energy of 13~TeV.
In Section~\ref{sec:total} we consider the results for the total 
cross section, and we discuss their scale dependence
and related scale uncertainties. In Section~\ref{sec:diff} 
we consider differential cross sections, and we perform
detailed comparisons between results in the pole and \ms schemes.
Finally, in Section~\ref{sec:running} we present new results 
obtained by using fixed and running values of the \ms mass,
and we compare them with the corresponding CMS data.
Our results are summarised in Section~\ref{sec:summa}.

\section{The heavy-quark cross section in the \texorpdfstring{$\overline{\text{MS}}$}{MSbar} scheme}
\label{sec:msbar}

Perturbative calculations of QCD scattering processes involve UV divergences. Part of the divergences are reabsorbed in the renormalisation of the QCD coupling $\as$, and the customary
procedure is to work in the \ms scheme. The removal of the UV divergences associated with quark masses also needs the choice of a renormalisation scheme. 
Different schemes lead to renormalised masses whose relative difference
is formally of ${\cal O}(\as)$. 
A `natural' renormalisation scheme is the pole scheme.
The renormalised quark mass in this scheme is determined order-by-order in the perturbative expansion by the pole of the
quark propagator, and, therefore, it corresponds to the customary meaning of mass for the case of a `physical' quark.
A possible alternative scheme is the use of \ms renormalisation also for the quark mass.

The top-quark mass in the pole scheme, \Mt, is related to the \ms mass at the scale \mum, $m_t(\mum)$, through the following perturbative relation:
\begin{equation}
\label{eq:polemsbar}
\Mt = m_t(\mum) \,d(m_t(\mum), \mum) =
m_t(\mum) \left(
1+\sum_{k=1}^\infty  \left(\frac{\as(\mum)}{\pi}\right)^k\, d^{(k)}(\mum)
\right).
\end{equation}
The first two perturbative coefficients $d^{(1)}$ and $d^{(2)}$ in Eq.~(\ref{eq:polemsbar}) have the values~\cite{Gray:1990yh,Fleischer:1998dw}
\begin{align}
  d^{(1)}(\mum) &= \frac{4}{3} + \lmnew\nonumber \,,
  \\[1ex]
  d^{(2)}(\mum) &= \frac{307}{32} + 2 \* \z2 + \frac{2}{3} \* \z2 \* \ln 2 - \frac{1}{6}\*\z3 
  + \frac{509}{72}\*\lmnew
  + \frac{47}{24}\*\lmnew^2 
  \nonumber\\
  &- \left( \frac{71}{144} + \frac{1}{3}\*\z2 + \frac{13}{36}\*\lmnew + \frac{1}{12}\*\lmnew^2 \right)\*n_f 
  \, ,
\label{eq:dcoef}
\end{align}
where
\begin{equation}
  \lmnew = 2 \ln(\mum/\mmunew) \,.
\end{equation}
The three-loop coefficient $d^{(3)}$ was computed in Refs.~\cite{Chetyrkin:1999qi,Melnikov:2000qh}, and the numerical result for $d^{(4)}$ was presented in Ref.~\cite{Marquard:2016dcn}.
Here and in the following, $\as(\mu)$ is the QCD coupling in the \ms renormalisation scheme, and its running with the scale $\mu$ is understood to be computed with $n_f=5$ light flavours. The light quarks are considered to be massless throughout the paper.

A main feature of the \ms mass is that it is a scale-dependent quantity, since it depends on the arbitrary mass renormalisation scale
\mum. An ensuing feature (which follows from Eq.~(\ref{eq:polemsbar}) and the fact that \Mt does not depend on \mum) is that the scale dependence of 
$m_t(\mum)$ is perturbatively computable. 
The dependence of $m_t(\mum)$ on the scale is driven by the renormalisation group equation
\begin{equation}
\label{eq:eveq} 
\f{d\ln m_t(\mum)}{d\ln \mumsq}=\gamma_m(\as(\mum)) \,,
\end{equation}
where the mass anomalous dimension $\gamma_m(\as)$ has the perturbative expansion
\begin{equation}
\label{eq:gamma}
\gamma_m(\as)=-\left(\gamma_0 \f{\as}{\pi}+{\cal O}(\as^2)\right) \,,
\end{equation}
and $\gamma_0=\frac{3}{4}C_F=1$. 
The perturbative expansion of the mass anomalous dimension in Eq.~(\ref{eq:gamma})
is explicitly known up to ${\cal O}(\as^4)$ 
\cite{Chetyrkin:1997dh,Vermaseren:1997fq}
and ${\cal O}(\as^5)$ \cite{Baikov:2014qja,Luthe:2016xec,Baikov:2017ujl}.
We note that comparing to the $\as$ evolution with $n_f=5$ flavours, the mass evolution is roughly a factor of two slower.
Indeed, the first coefficient $\beta_0$ of the QCD $\beta$ function is
$\beta_0=\frac{11 C_A}{12} - \frac{n_f}{6}=\frac{33}{12} - \frac{n_f}{6}$, and, setting $n_f=5$ we have
\begin{equation}
\label{eq:slow}
  \gamma_0=\f{12}{23}\, \beta_0\, . 
\end{equation}

The actual value of $m_t(\mum)$ at any scale \mum can be specified by relying on the knowledge of \Mt and, therefore, by using Eq.~(\ref{eq:polemsbar}).
Alternatively, $m_t(\mum)$ can be obtained by solving its renormalisation group equation, provided the \ms mass is known at some reference scale. The choice
of the reference scale is arbitrary, and it can be done independently of the knowledge of \Mt. For instance, one can simply choose a specific value of 
\mum (e.g., analogously to the choice of the mass of the $Z$ boson as a typical reference scale for the running coupling $\as$). The reference scale can also be chosen
through an `intrinsic' definition, as a scale of the same order as the \ms mass itself. A customary intrinsic definition corresponds to the scale 
$\mbar$ such that $m_t(\mbar)= \mbar$. In this case the coefficients 
$d^{(1)}(\mbar)$ and $d^{(2)}(\mbar)$ in Eq.~(\ref{eq:dcoef}) are positive and of order unity,
and, therefore, the reference mass $\mbar$ is typically smaller than the pole mass \Mt by about 10~GeV.

We note that the reference scale $\mbar$ has no special physical meaning, since an intrinsic \ms reference scale can be introduced differently. For instance, one can define the scale $\mbarlam$ such that 
$m_t(\lambda\, \mbarlam)= \mbarlam$, where $\lambda$ is a parameter of order unity. The choice $\lambda=1$ leads to $\mbarlam=\mbar$. Different choices
of $\lambda$ lead to scales $\mbarlam$ whose relative difference from $\mbar$
is of ${\cal O}(\as)$.

We also recall that the features of the \ms scheme are unchanged by using any other 
\ms-like scheme that is obtained by a perturbative redefinition of $m_t(\mum)$
with scale-independent coefficients. The relative difference between the 
running masses of the two schemes is of ${\cal O}(\as)$ 
(this affects Eq.~(\ref{eq:polemsbar}) starting from its first-order coefficient
$d^{(1)}$), and the first-order coefficient $\gamma_0$ of Eq.~(\ref{eq:gamma})
is unchanged (the change of scheme affects Eq.~(\ref{eq:gamma}) starting from
${\cal O}(\as^2)$).

It is well known that the renormalised pole mass is affected by a renormalon 
ambiguity~\cite{Beneke:1994sw,Bigi:1994em,Beneke:1994rs}. More precisely,
at large values of $k$ the perturbative coefficients $d^{(k)}$ in Eq.~(\ref{eq:polemsbar}) are factorially 
growing with $k$.
This implies that a non-perturbative ambiguity of ${\cal O}(\Lambda_{\rm QCD})$ affects the definition of the pole mass itself 
(no corrections of ${\cal O}(\Lambda_{\rm QCD})$ affect the \ms mass).
The knowledge of the coefficients $d^{(k)}$ with $k \leq 4$ can be combined with that
of the asymptotic factorial behavior~\cite{Beneke:1994rs} to obtain 
approximations~\cite{Ayala:2014yxa, Beneke:2016cbu, Hoang:2017btd}
of the perturbative coefficients of Eq.~(\ref{eq:polemsbar})
beyond the four-loop order. High-order approximations of Eq.~(\ref{eq:polemsbar})
can also be used to estimate
the renormalon ambiguity on the value of the pole mass.
In Ref.~\cite{Beneke:2016cbu} the ambiguity is estimated to be about $110$~MeV, while Ref.~\cite{Hoang:2017btd} estimates it to be about 250~MeV.
We note that both estimates are of the order or below the accuracy that can be reasonably achieved in LHC measurements
of the top-quark mass.

The renormalon ambiguity does not only plague the pole mass definition, but it can in principle affect the perturbative expansion of \ttb cross sections.
The total cross section expressed in terms of the \ms mass is expected to be not affected by renormalons at ${\cal O}(\Lambda_{\rm QCD})$.
However, the \ttb total cross section is not directly measurable in practice.
Renormalon effects are unavoidable~\cite{FerrarioRavasio:2018ubr} in the perturbative computation of \ttb production observables that are defined by 
realistic selection cuts applied in LHC experiments.
Even if LHC experimental measurements are often extrapolated to the full phase space,
the extrapolation procedure is not able to (theoretically) correct for renormalon effects.

At the LHC top-quark physics can be studied either indirectly or directly.
By indirectly we mean through quantities in which the top quark appears as a virtual
(off-shell) particle that is not directly observed. For instance, since the Higgs boson is mostly produced by gluon fusion through the coupling to a top-quark loop,
studies of Higgs boson production give information on the top quark, including the effects of its mass. For all the quantities in which the top quark appears indirectly,
mass renormalisation can be carried out equivalently in any renormalisation scheme.
For instance, one can introduce \ms renormalisation and the \ms mass without using
and even mentioning the pole mass (one can also do the opposite, of course).
For all these quantities the \ms and pole masses can be introduced on equal footing.

The direct studies of top-quark physics at the LHC are those in which the top 
quark (and/or antiquark) or, more precisely, its decay products, are directly observed in the final state. These studies are based on a definite physical picture that fully relies 
on the concepts of pole mass, \Mt, and width, $\Gamma_t$, of the top quark.
The top quark is so heavy ($\Mt \sim 173$~GeV from direct measurements at the Tevatron
and the LHC~\cite{Tanabashi:2018oca}) and so unstable 
($\Gamma_t \sim 1.4$~GeV~\cite{Tanabashi:2018oca}) that it decays by weak interactions before strong interactions come into play and form bound states. Therefore, the top quark is viewed as a `physical', though unstable, particle (its pole mass and width
gain a physical meaning) that manifests itself in a resonance peak in physical cross sections.
LHC experimental data on top-quark production entirely rely on this physical picture,
since the top-quark signal is extracted by the quasi-resonant behaviour of the (supposed) decay products of the top quark.
No LHC experimental data on top-quark production can be obtained without referring
to the pole of the propagator of the top quark
(i.e., without having introduced the concept of pole mass).
Therefore, the pole mass and the \ms mass do not appear on equal footing in the context of top-quark production at the LHC. The pole mass has a primary role, and the \ms mass has (somehow) an auxiliary role.
Note that this is not only a conceptual aspect, since the difference between \Mt
and $m_t(\mum)$ (at scales \mum of the order of the top-quark mass) can be as large as about 10~GeV, and, hence, much larger than $\Gamma_t$. Therefore,
the pole and \ms masses cannot be regarded as being approximately equal for practical experimental purposes. We also note that the size of the renormalon effects on \Mt
is definitely smaller than $\Gamma_t$ and, hence, renormalons do not change the picture of the top quark as a `physical' unstable particle. The main caveat to this picture is due to the fact that the top quark carries colour charge, and QCD colour is confined through hadronization and not observable in the final state. This implies that colour confinement produces quantitative corrections on the identification 
of the decay products of the top quark. However, such effects are not related to the difference between the pole and \ms masses.

The experimental treatment of the top quark as a physical unstable particle has a direct correspondence in theoretical calculations based on the narrow-width approximation. In the limit $\Gamma_t \ll \Mt$, top-quark production is computed
by setting the top quark on shell and with a mass equal to the pole mass 
\Mt.\footnote{The top-quark decay can also be computed and included within the same approximation. Throughout this paper we do not consider the decay of the top quark.}
We use the shorthand notation $\sigma(\Mt; X)$ to generically denote cross sections
or differential cross sections for on-shell \ttb 
production.\footnote{The variable $X$ in $\sigma(\Mt; X)$ can directly refer to differential cross sections, $d\sigma/dX$, or it can generically refer to a set of acceptance cuts that specify fiducial cross sections. The variable $X$ is absent in the case of the \ttb total cross section.} The perturbative QCD calculation of $\sigma(\Mt; X)$ defines the perturbative function 
$\sigma(\as(\muR), \muR,\muF; \Mt; X)$ that is computed order-by-order as a series expansion in powers of $\as(\muR)$. The dependence on the auxiliary remormalization scale \muR and factorisation scale \muF is due to the \ms renormalisation of 
$\as$ and the \ms factorisation of the parton distribution functions (PDFs) of the colliding hadrons. The perturbative expansion 
of 
$\sigma(\as(\muR), \muR,\muF; \Mt; X)$ up to NNLO can be written as
\begin{equation}
  \label{eq:hadro-mpole}
  \sigma_\text{NNLO}(\as(\muR), \muR,\muF; \Mt; X) = 
     \sum_{i=0}^2\, \left(\frac{\as(\muR)}{\pi}\right)^{i+2}\, \sigma^{(i)}(\Mt;\muR,\muF; X) 
  \; ,
\end{equation}
where the perturbative coefficients $\sigma^{(i)}(\Mt;\muR,\muF; X)$ explicitly depend on the pole mass \Mt of the on-shell top quark and antiquark that are produced in the final state.

In the context of Eq.~(\ref{eq:hadro-mpole}) (and its generalization to higher
perturbative orders), we remark on the fact that \Mt is not simply a parameter of the QCD Lagrangian, but it is also, and importantly, a key kinematical parameter of the phase space. The final-state top quark has a mass \Mt, and the on-shell constraint $p^2=\Mt^2$ affects each of the components $p^\nu$ of the four momentum
$p$ of the top quark. This in turn produces a dependence on \Mt of all the kinematical variables of the produced final state. For instance, if the kinematical variable $X$ is the invariant mass \mtt of the \ttb pair, it has an implicit dependence on \Mt, which in particular leads to the constraint
$\mtt> 2\Mt$. Therefore, the differential cross section with respect to 
\mtt has a `physical' threshold at $\mtt=2\Mt$, and it vanishes for smaller values of \mtt (this is true for the perturbative cross sections
$\sigma^{(i)}$ in Eq.~(\ref{eq:hadro-mpole}) at each perturbative order). Obviously, the kinematical/phase space
dependence of the cross section on \Mt is the consequence of the underlying dynamical approximation, namely, the fact that the cross section in 
Eq.~(\ref{eq:hadro-mpole}) deals with the production of 
an `unstable' 
top quark with pole mass \Mt in the limit of vanishing width $\Gamma_t$.

Throughout the paper we refer to the cross section in Eq.~(\ref{eq:hadro-mpole})
(and its generalization to higher orders) as the on-shell cross section computed in the pole mass scheme.

Starting from the pole scheme cross section $\sigma(\as(\muR), \muR,\muF; \Mt; X)$,
we theoretically define the \ms cross section $\bar\sigma$ through a formal replacement of \Mt with its dependence on the \ms mass $m_t(\mum)$. Our definition is
\begin{equation}
\label{eq:all}
\bar\sigma(\as(\muR), \muR,\muF; \mum, m_t(\mum); X)
= \sigma(\as(\muR), \muR,\muF; \Mt=m_t(\mum)\,d(m_t(\mum),\mum) ; X) \;,
\end{equation}
where the right-hand side of the equation is the pole scheme cross section in which the pole mass \Mt has been expressed in terms of the \ms mass through the all-order relation in Eq.~(\ref{eq:polemsbar}).
The cross sections $\sigma$ and $\bar\sigma$ in Eq.~(\ref{eq:all}) are equal if regarded as formal expressions to all orders in $\as$. The order-by-order expansions of
$\sigma$ and $\bar\sigma$ are instead different.
The perturbative expansion of $\bar\sigma$ is simply obtained by considering the 
right-hand side of Eq.~(\ref{eq:all}) and expressing $d(m_t(\mum),\mum)$
as a function of $\as(\muR)$ and $m_t(\mum)$ (see Eq.~(\ref{eq:polemsbar})).
The expression, $\bar\sigma_\text{NNLO}$, of the \ms scheme cross section up to NNLO
is 
\begin{equation}
  \label{eq:xs_MSbar}
  \bar\sigma_\text{NNLO}(\as(\muR), \muR,\muF;\mum, m_t(\mum) ; X) = 
    \sum_{i=0}^2 \! \left(\frac{\as(\muR)}{\pi}\right)^{i+2} \bar\sigma^{(i)}(m_t(\mum);\mum,\muR,\muF; X) 
  \, ,
\end{equation}
and the explicit contributions $\bar\sigma^{(i)}$ are
\begin{eqnarray}
 \label{eq:barlo}
&& \!\!\!\!\!\!\!\!
\bar\sigma^{(0)}(m_t(\mum);\muF; X) = \bigg[\sigma^{(0)}(m;\muF; X)\bigg]_{m=m_t(\mum)}
\\[3ex]
 \label{eq:barnlo}
&& \!\!\!\!\!\!\!\!
\bar\sigma^{(1)}(m_t(\mum);\mum,\muR,\muF; X)
= \bigg[ \sigma^{(1)}(m;\muR,\muF; X) 
+ d^{(1)}(\mum)\,m \,\partial_{m} \sigma^{(0)}(m;\muF; X)
\bigg]_{m=m_t(\mum)} 
\end{eqnarray}
\begin{eqnarray}
 \label{eq:barnnlo}
&& \!\!\!\!\!\!\!\!
\bar\sigma^{(2)}(m_t(\mum);\mum,\muR,\muF; X) =
   \Bigg[ \sigma^{(2)}(m;\muR,\muF; X) \nn \\
   && \;\;\;\;\;\; + \; m \Bigg(
   d^{(1)}(\mum) \,\partial_{m} \sigma^{(1)}(m;\muR,\muF; X) 
   + \frac{1}{2} \left(d^{(1)}(\mum)\right)^2 \,m\,
       \partial^2_{m} \sigma^{(0)}(m;\muF;X) \\
   && \;\;\;\;\;\; + \,d^{(2)}(\mum) \,\partial_{m} \sigma^{(0)}(m;\muF; X)
   + \beta_0 \, d^{(1)}(\mum) \ln\left(\frac{\muRsq}{\mumsq} \right)     
     \partial_{m} \sigma^{(0)}(m;\muF;X)
   \Bigg)
   \Bigg]_{m=m_t(\mum)} \;,\nn
\end{eqnarray}
where $d^{(1)}(\mum)$ and $d^{(2)}(\mum)$ are the coefficients in 
Eq.~(\ref{eq:dcoef}), and the term proportional to $\beta_0$
in Eq.~(\ref{eq:barnnlo}) arises from expressing Eq.~(\ref{eq:polemsbar})
in terms of $\as(\muR)$ (rather than $\as(\mum)$).

We comment on some main features of the definition of the cross section $\bar\sigma$
in the \ms scheme.

The perturbative contributions $\bar\sigma^{(i)}$ are given in terms of the corresponding contributions $\sigma^{(i)}$ as computed for on-shell \ttb production
in the pole mass scheme. The cross sections $\bar\sigma^{(i)}$ depend on the on-shell cross sections
$\sigma^{(i)}$ and their derivatives ($\partial_m\equiv \partial/\partial m$) with respect to the top-quark mass. These are partial derivatives at fixed values of the auxiliary scales $\muR, \muF$ and \textit{fixed} values of $X$.

The \ttb total cross section only depends on the top-quark mass (and auxiliary scales). Fiducial and differential cross sections depend on the additional 
variables $X$.
As previously discussed, in the perturbative calculation for on-shell top quarks, the variables $X$ are kinematically correlated to the pole mass. The formal 
definition of the \ms cross section in Eqs.~(\ref{eq:all})--(\ref{eq:barnnlo}) produces ensuing kinematical correlations between the variables $X$ and the \ms mass $m_t(\mum)$.

We also note that the partial derivatives $\partial^k \ln \sigma^{(i)}/(\partial \ln m)^k$ can be very sizeable in some cases. If this happens, the fixed-order expansion
of the \ms scheme cross section can become quantitatively unstable, although the all-order equality in Eq.~(\ref{eq:all}) remains valid.

The perturbative contributions $\bar\sigma^{(i)}$ to the \ms cross sections depend on the \ms mass $m_t(\mum)$ and on the mass renormalisation scale \mum.
The auxiliary scale \mum is formally arbitrary. The usual procedure to obtain quantitative predictions in the presence of auxiliary scales, such as $\muR, \muF$ and \mum, is to assign them some central values and, then, consider variations around these central values. There are no reasons to consider perturbative predictions by uniquely fixing the value of \mum (e.g., by setting $\mum=\mbar$) without examining the effect of varying \mum.

As a final general comment on the \ms scheme cross section $\bar\sigma$ we note that,
analogously to the pole scheme cross section, it regards top-quark production in the limit of vanishing top width $\Gamma_t$. The formal definition of $\bar\sigma$ in 
Eq.~(\ref{eq:all}) does not introduce any physical corrections due to the finite width
of the top quark.

In the following, we comment on the actual structure of the perturbative coefficients in 
Eqs.~(\ref{eq:xs_MSbar})--(\ref{eq:barnnlo}).
At the LO there are no mass renormalisation effects to be considered. The LO
\ms scheme cross section $\bar\sigma^{(0)}$ in Eq.~(\ref{eq:barlo}) is equal to
the on-shell cross section $\sigma^{(0)}$ in Eq.~(\ref{eq:hadro-mpole}), apart from the replacement of the pole mass \Mt with the \ms mass $m_t(\mum)$.
Since we are dealing with the production of on-shell quarks, this formal
replacement is questionable from a physics viewpoint. For instance, the question becomes evident by considering the differential cross sections with respect to  
\mtt. The pole scheme differential cross section $d\sigma/d\mtt$
(at any perturbative order) has a physical threshold at $\mtt= 2\Mt$, whereas
in the \ms cross section at LO the threshold is at $\mtt= 2m_t(\mum)$.
If the difference between \Mt and $m_t(\mum)$ is much larger than the width
$\Gamma_t$, such displacement of the production threshold is definitely unphysical.

At the NLO, one-loop corrections on internal quark lines require mass 
renormalisation, which can be carried out either in the pole or the \ms scheme.
However, the NLO cross section $\sigma^{(1)}$ in Eq.~(\ref{eq:hadro-mpole}) still involves on-shell top quarks with pole mass \Mt. The NLO \ms cross section 
$\bar\sigma^{(1)}$ in Eq.~(\ref{eq:barnlo}) includes two contributions: one contribution is simply $\sigma^{(1)}$ with the replacement $\Mt \to m_t(\mum)$,
and the other contribution is controlled by the mass derivative of the LO on-shell cross section $\sigma^{(0)}$. This second contribution represents the correction that is applied to $\bar\sigma$ for having naively identified \Mt with $m_t(\mum)$
at the LO. For instance, the correction is quantitatively very large for the differential cross section $d\bar\sigma^{(1)}/d\mtt$ close to the threshold region where $\mtt \sim 2m_t(\mum) \sim 2\Mt$, since the mass derivative of 
$d\bar\sigma^{(0)}/d\mtt$ is very large in this region.

A similar discussion can be extended at NNLO and higher orders. In particular, the mass derivatives of the on-shell cross sections $\sigma^{(j)}$ that appear in the perturbative contributions $\bar\sigma^{(i)}$ $(j < i)$ partly originate from having replaced the pole mass \Mt with the \ms mass $m_t(\mum)$ in the lower-order cross sections for on-shell top-quark production.

We summarise the main points of our general discussion of the \ms scheme cross section. Dealing with the production of on-shell top quarks with pole mass \Mt,
we have considered the pole scheme cross section $\sigma(\as, \Mt, X)$ in 
Eq.~(\ref{eq:hadro-mpole}). The \ms scheme cross section 
$\bar\sigma(\as, m_t(\mum), X)$ is then introduced through the pole scheme cross section and the mass relation in Eq.~(\ref{eq:polemsbar}). The cross section 
$\bar\sigma$ is defined by using the formal all-order identity in Eq.~(\ref{eq:all}).
In this relation, the cross section variables $X$ (which can depend on the on-shell kinematics of the produced top quarks) are considered to be independent of the pole mass
\Mt. The cross sections $\sigma$ and $\bar\sigma$ differ order-by-order in $\as$.
Owing to the perturbative nature of the definition in Eq.~(\ref{eq:all}) and using mass scales \mum of the order of the top-quark mass (so that the coefficients
$d^{(k)}(\mum)$ in Eq.~(\ref{eq:polemsbar}) are of order unity), we expect the following perturbative behaviour. At low perturbative orders, $\sigma$ and 
$\bar\sigma$ can give quantitative results that are consistent (within perturbative uncertainties), and the difference can be larger for cross sections that depend on kinematical variables $X$ that are more sensitive to kinematic thresholds related to on-shell top-quark production. At higher perturbative orders, 
$\sigma$ and $\bar\sigma$ can give very similar quantitative results, thus leading to an equivalent perturbative description. As we briefly discuss below, at such perturbative orders, the \ms formulation can take advantage of the dynamical features of the running mass $m_t(\mum)$.

In the case of cross sections that depend on physical scales $X$ that are much larger than the mass of the top quark (e.g., if $X=\mtt$ at high values of \mtt), 
\ttb production takes place in a multiscale dynamical regime. Therefore, we can expect that an improved perturbative description can be achieved in the context of the \ms scheme by using a running mass $m_t(\mum)$ with a dynamical value of the renormalisation scale \mum. We also note that in such high-scale regime the variable $X$ 
has a weaker dependence on the pole mass of the produced on-shell top quark.

In the next Section we present detailed quantitative studies of the perturbative features of the \ms scheme cross sections. We also discuss an implementation of the running mass $m_t(\mum)$ in the computation of the differential cross section with respect to the \ttb invariant mass \mtt.

\section{Results}
\label{sec:resu}

In this Section we present inclusive and differential results for \ttb production in the \ms scheme up to NNLO, and we compare them with the corresponding results obtained in the pole scheme.
All our results are based on the calculation of \ttb production up to NNLO that is reported in Refs.~\cite{Bonciani:2015sha,Catani:2019iny,Catani:2019hip}.
The calculation is carried out within the \Matrix~\cite{Grazzini:2017mhc} framework,
which features a completely automated implementation of the $q_T$ subtraction formalism~\cite{Catani:2007vq} at the NNLO. The core of the \Matrix framework is the Monte Carlo program \Munich\footnote{\Munich is the 
abbreviation of ``MUlti-chaNnel Integrator at Swiss~(CH) precision'' --- an automated parton-level
NLO generator by S.~Kallweit.}, which includes a fully automated implementation of the
NLO dipole subtraction method for massless~\cite{Catani:1996jh,Catani:1996vz}
and massive~\cite{Catani:2002hc} partons,
and an efficient phase space integration.
All the required (spin- and colour-correlated) tree-level and one-loop (squared) amplitudes
are obtained by using \OpenLoops~\cite{Cascioli:2011va,Buccioni:2017yxi,Buccioni:2019sur}. The required two-loop amplitudes are available in a numerical form~\cite{Baernreuther:2013caa}.
More details on the implementation of \ttb production in \Matrix can be found in Ref.~\cite{Catani:2019hip}.

The calculation of Ref.~\cite{Catani:2019hip} directly leads to the numerical evaluation of the perturbative contributions $\sigma^{(0)}$, $\sigma^{(1)}$
and $\sigma^{(2)}$ (see Eq.~(\ref{eq:hadro-mpole})) to the cross sections for on-shell \ttb production in the pole scheme.
According to Eqs.~(\ref{eq:xs_MSbar})--(\ref{eq:barnnlo}), the calculation of the \ms cross section $\bar\sigma$ up to NNLO requires the computation of the first and second derivatives of the LO result $\sigma^{(0)}$ with respect to the mass,
and of the first derivative of the NLO corrections $\sigma^{(1)}$, computed in the pole scheme.
This calculation is performed by computing the cross sections $\sigma^{(0)}$ and $\sigma^{(1)}$ for several values of the top-quark mass around $m=m_t(\mum)$, and performing a quadratic fit of the results, from which the numerical values of the relevant derivatives are obtained. This procedure is carried out for the total cross section and, analogously, for each bin in the variable $X$ of the considered differential distributions $d\sigma/dX$.

To present our quantitative results, we focus on $pp$ collisions at the 
centre-of-mass energy $\sqrt{s}=13$~TeV.
In Section \ref{sec:total} and \ref{sec:diff}
we consider perturbative calculations in the pole scheme and in the \ms scheme by using values of the renormalisation scale \mum
of the order of the top-quark mass.
We use $n_f = 5$ massless quark flavours and the corresponding NNPDF31 sets~\cite{Ball:2017nwa} of parton distribution functions (PDFs) with $\as(m_Z)=0.118$. In particular, N$^n$LO (with $n = 0,1,2$)
predictions are obtained by using PDFs at the corresponding perturbative order and the evolution of $\as(\muR)$ at $(n + 1)$-loop order, as provided by the PDF set.
Our results in the pole scheme with $\Mt = 173.3\text{ GeV}$~\cite{Tanabashi:2018oca}
are compared with the corresponding results in the \ms scheme with
$\mbar = 163.7\text{ GeV}$.
These two values of \Mt and $\mbar$
are numerically related by mass renormalisation 
at the NNLO,
namely, we use the relation in Eq.~(\ref{eq:polemsbar}) at three-loop order 
by including the coefficients
$d^{(1)}$, $d^{(2)}$ and $d^{(3)}$~\cite{Gray:1990yh,Fleischer:1998dw,Chetyrkin:1999qi,Melnikov:2000qh}
(we set $d^{(k)}=0$ if $k \geq 4$).
We note that the same value of $\mbar$ is used regardless of the order of the calculation.

Unless otherwise stated, the top-quark mass (either $\mbar$ or \Mt) is used in our calculations as the central value $\mu_0$ for the renormalisation (both \muR and \mum) and factorisation scales.
We use the customary procedure of performing scale variations around the central scales
to estimate the uncertainties from perturbative contributions at higher orders or,
more precisely, to roughly set a lower limit on such uncertainties.
The scale uncertainty bands for the predictions in the pole scheme are obtained by setting $\mu_0=\Mt$ and performing independent variations of \muR and \muF. 
We set 
$\mu_i = \xi_i \mu_0$, 
and we vary the parameter $\xi_i$ according to $\xi_i =\{ 1/2,1,2 \}$ with the constraints $\mu_i/\mu_j \leq 2$  ($i,j = R,F$). This prescription leads to the customary 7-point scale uncertainty.
In the case of the \ms scheme, we have an additional auxiliary scale, \mum, which
(as discussed in Section~\ref{sec:msbar}) has to be varied.
We perform an independent variation of the three auxiliary scales, by setting
$\mu_i = \xi_i \mu_0$ (here $\mu_0=\mbar$) and varying $\xi_i$  as $\xi_i = \{ 1/2,1,2\}$, with the constraints $\mu_i/\mu_j \leq 2$ ($i,j = R,F,m$).
This prescription leads to a 15-point scale variation.
By varying \mum in the interval $0.5\,\mbar < \mum < 2\,\mbar$, the \ms mass varies in the range $155.5\text{~GeV}\ltap m_t(\mum) \ltap 173.3\text{~GeV}$. This dependence of $m_t(\mum)$ on \mum is computed at NNLO accuracy\footnote{We have checked that the variation range of $m_t(\mum)$ is almost unchanged by using the evolution equation (\ref{eq:eveq}) at lower perturbative orders.}
(i.e., we consider the evolution equation (\ref{eq:eveq}) with 
the anomalous dimension $\gamma_m(\as)$ of Eq.~(\ref{eq:gamma}) that is evaluated up
to ${\cal O}(\as^3)$) by using the package \CRunDec~\cite{Schmidt:2012az}.
We note that the upper limit on $m_t(\mum)$ is
very close to the value of the pole mass \Mt.

\subsection{Total cross section}
\label{sec:total}

We start the presentation of our results by considering the \ttb total cross section. In Table~\ref{table:totalXS} we compare the results at LO, NLO and NNLO 
in the pole scheme with the corresponding results in the \ms scheme.
The scale uncertainties of the \ms scheme results as evaluated in different ways
(see the comments below) are also presented.
We have checked that the \ms scheme results at fixed $\mum=\mbar$, including \muR and \muF variations around $\mu_0=\mbar$ (third column of Table~\ref{table:totalXS}), are in excellent quantitative agreement with those obtained by using the numerical program \Hathor~\cite{Aliev:2010zk, Langenfeld:2009wd}.

{\renewcommand{\arraystretch}{1.6}
\begin{table}[t]
\begin{center}
\begin{tabular}{l|C{.16\textwidth}|C{.16\textwidth}|C{.12\textwidth}|C{.12\textwidth}|C{.12\textwidth}}
\hline
scheme & pole & \multicolumn{4}{c}{\ms} \\
\hline
variation & 7-point & 15-point & $\mu_m = \mu_0$ & $\mu_{R/F} = \mu_0$ & $\mu_{R/F} = \mu_m$ \\
 \hline
 LO (pb)  & $478.9\;^{+29.6\%}_{-21.4\%}$ & $625.7\;^{+29.4\%}_{-21.9\%}$ & $^{+29.4\%}_{-21.3\%}$ & $^{+24.7\%}_{-21.9\%}$  & $^{+1.5\%}_{-1.5\%}$ \\
 NLO (pb) & $726.9\;^{+11.7\%}_{-11.9\%}$ & $826.4\;^{+7.6\%}_{-9.7\%}$ & $^{+7.6\%}_{-9.6\%}$ & $^{+5.6\%}_{-9.7\%}$  & $^{+1.2\%}_{-1.2\%}$ \\
 NNLO (pb)& $794.0\;^{+3.5\%}_{-5.7\%}$ & $833.8\;^{+0.5\%}_{-3.1\%}$ & $^{+0.4\%}_{-2.9\%}$ & $^{+0.3\%}_{-3.1\%}$  & $^{+0.0\%}_{-0.3\%}$ \\ \hline
\end{tabular}
\end{center}
\vspace*{-0.6cm}
\caption{
Total cross section at $\sqrt{s}=13$~TeV in the pole scheme with $M_t = 173.3\text{ GeV}$  and in the \ms scheme with $\mbar = 163.7\text{ GeV}$. 
The central results refer to the scales $\mu_R=\mu_F=\mu_0=M_t$ in the pole scheme and
$\mu_R=\mu_F=\mu_m=\mu_0=\mbar$ in the \ms scheme. The scale dependence is computed by performing independent scale variations by a factor of two around the central scales,
as described in the text. In the case of the \ms scheme,
the uncertainties obtained with different prescriptions for the scale variations are also shown.
}
\label{table:totalXS}
\end{table}
}

We present some comments on the scale dependence of the results in Table~\ref{table:totalXS}.
We observe that the scale uncertainties obtained in the \ms scheme by
using the 15-point scale variation, by keeping \mum fixed or by keeping \muR and \muF fixed (see, correspondingly, the second, third or fourth column in Table~\ref{table:totalXS})
are quantitatively very similar.
The reason for this similarity is that the total cross section increases as \mum increases, while it has the opposite dependence on \muR and \muF.
Indeed, if \mum increases, the value of $m_t(\mum)$ decreases, thereby leading to an increase of the cross section.
On the contrary, 
the cross section decreases as \muR increases, and it slightly decreases as \muF increases. This \muR dependence is due to the overall proportionality of the cross section to the factor $\as^2(\muR)$ (see Eq.~(\ref{eq:xs_MSbar})). The \muF dependence is due the fact that the \ttb cross section is sensitive to relatively large momentum fractions of the colliding partons, and in this kinematical region 
the scaling violations of the PDFs are slightly negative.
Moreover, we note that the absolute variation of the cross section that is obtained with the \mum scan is similar in size to the one obtained by varying \muR (whose dependence dominates in the 7-point variation of \muR and \muF).
These features imply that the uncertainty obtained by varying all the three scales simultaneously ($\muR=\muF=\mum$) is much smaller than the one obtained with independent variations.
We also note that larger scale uncertainties are obtained if the constraint $1/2 \leq \mum/\mu_i \leq 2$, with $i = R,F$, is relaxed (the corresponding results are not shown in Table~\ref{table:totalXS}). This fact is also in line to the observations made above.

Using Table~\ref{table:totalXS}, we can compare the results in the pole and \ms schemes,
and we note the following main features. Order-by-order in perturbation theory, the total cross section at central scales is larger in the \ms scheme: the increase is about 30\% at LO, 15\% at NLO and 5\% at NNLO. In particular, the NNLO results in the two schemes are very similar. Referring to the full scale dependence
(as given in the first two columns of Table~\ref{table:totalXS}), the order-by-order 
differences between the two schemes are comparable to the corresponding scale dependence effects. Therefore, at each perturbative order the predictions in the \ms scheme are consistent with the corresponding predictions in the pole scheme within their scale uncertainties. At LO the results in the two schemes have a very similar scale dependence. At higher orders the \ms scheme results have a reduced scale dependence. The reduction of the scale dependence is moderate at NLO and more sizeable at NNLO. 
We note, however, that the NNLO scale uncertainties of the \ms scheme result are highly asymmetric: the upward variation is much smaller than the downward variation.
This means that the \ms scheme cross section is computed close to a local maximum
(namely, close to a region of
local minimal sensitivity~\cite{Stevenson:1981vj}) of the scale dependence of the NNLO result. This may in turn lead to an underestimate of the perturbative uncertainty due to
higher-order corrections.

In Ref.~\cite{Langenfeld:2009wd} it was pointed out that the perturbative convergence of the total cross section in the \ms scheme appears to be faster than in the pole scheme. Such behaviour is indeed visible by considering the results in
Table~\ref{table:totalXS} at central values of the scales. To quantify the effect we can introduce the $K$-factors, $K_\text{(N)NLO} = \sigma_\text{(N)NLO}/\sigma_\text{(N)LO}$, namely
the ratios of the cross section results at two subsequent orders. In the pole scheme
we have $K_\text{NLO}=1.52$ and $K_\text{NNLO}=1.09$,
while in the \ms scheme we have $K_\text{NLO}=1.32$ and $K_\text{NNLO}=1.01$.
At each perturbative order the $K$-factor in the \ms scheme is smaller than that in the pole scheme, thus leading to a faster apparent convergence of the perturbative expansion.
We recall that we are using an NNLO relation to obtain the value of $\mbar$ from
$\Mt = 173.3\text{ GeV}$.
The faster apparent convergence of the \ms scheme results would be partly reduced by considering a strictly formal order-by-order comparisons between the two mass schemes. Such comparison implies the use of Eq.~(\ref{eq:polemsbar}) 
at each corresponding order, and it leads to $\mbar = 165.8\text{ GeV}$ at LO
and $\mbar=164.2\text{ GeV}$ at NLO.
The corresponding LO and NLO cross sections in the \ms scheme have the values
$589.0$ pb and $808.6$ pb, respectively. Therefore, in this case the $K$-factors
in the \ms scheme would be $K_\text{NLO}=1.37$ and $K_\text{NNLO}=1.03$.

We are not able to offer a physical interpretation of the faster apparent convergence of the \ms scheme results, but we do have a technical explanation for it. The explanation uses the structure of the pole and \ms scheme cross sections in 
Eqs.~(\ref{eq:hadro-mpole})--(\ref{eq:barnnlo}), and it is discussed below.

We first consider the NLO $K$-factor. The LO cross section  $\bar \sigma^{(0)}$
in the \ms scheme is simply obtained by evaluating the LO cross section 
$\sigma^{(0)}$ in the pole scheme with a value of the top-quark mass ($\mbar = 163.7\text{ GeV}$) that is significantly lower than the value of \Mt. Since the on-shell 
total cross section is a decreasing function of the top-quark mass, the LO cross section increases in going from the pole scheme to the \ms scheme. Such increase
\textit{partly} contributes to a decrease of the value of $K_\text{NLO}$. A further decrease of 
$K_\text{NLO}$ is produced by the NLO radiative corrections. The NLO correction
$\sigma^{(1)}$ in the pole scheme is positive and sizeable. The NLO correction
$\bar \sigma^{(1)}$ in the \ms scheme receives the additional contribution of
$d^{(1)} m \partial_m \sigma^{(0)}$ (see Eq.~(\ref{eq:barnlo})) that is
\textit{negative} ($d^{(1)}$ is positive, and the derivative $\partial_m \sigma^{(0)}$
is negative since $\sigma^{(0)}$ is a decreasing function of the top-quark mass) and not small.
Therefore, we have $\bar \sigma^{(1)} < \sigma^{(1)}$, and this effect also decreases the size of $K_\text{NLO}$ in going from the pole scheme to the \ms scheme.

The discussion of the NNLO $K$-factor follows the same lines as at NLO. We simply notice two main effects in going from the pole scheme to the \ms scheme: the NLO cross section increases and the NNLO correction decreases 
(because in Eq.~(\ref{eq:barnnlo}) the coefficients $d^{(1)}$ and $d^{(2)}$
are positive and the mass derivatives of the cross sections are negative). Both effects
contribute to decrease the value of $K_\text{NNLO}$, thus producing a faster apparent convergence. Moreover, we point out that both effects are the consequence of two basics facts: the pole scheme cross section is a decreasing function of the top-quark mass with radiative corrections that are relatively large at lower orders; the top-quark mass $\mbar$ that is used in the \ms scheme calculation is significantly lower than \Mt. Therefore, the features of faster apparent convergence of the \ttb
cross section with respect to the behaviour in the pole scheme are common to any renormalisation scheme that perturbatively introduces a renormalisation mass that is systematically smaller than \Mt.

From the above discussion one is led to conclude that the computation in the \ms scheme leads to an improved perturbative stability with respect to the computation in the pole scheme. However, we find that such pattern of perturbative convergence is strongly dependent on the choice of the central values of the auxiliary scales.
This fact can be observed, for instance, in Table~\ref{table:scales} where we report
the results of Table~\ref{table:totalXS} at central scales and two additional sets of results obtained in the pole and \ms schemes by using different central values of the auxiliary scales.

{\renewcommand{\arraystretch}{1.6}
\begin{table}[t]
\begin{center}
\begin{tabular}{l|C{.16\textwidth}|C{.16\textwidth}|C{.16\textwidth}|C{.16\textwidth}}
\hline
scheme & pole & \ms & \ms & pole  \\
\hline
\\[-5ex]
\multirow{2}{*}[-0.3ex]{central scale choice} & \multirow{2}{*}{$\mu_{R/F}=M_t$} & $\mu_{R/F}=\mbar$ & $\mu_{R/F}=\mbar$ & \multirow{ 2}{*}{$\mu_{R/F}=M_t/2$}\\[-1ex]
& & $\mu_m=\mbar/2$ & $\mu_m=\mbar$ & \\
 \hline
 LO (pb)  & $478.9$ & $488.9$ & $625.7$ & $619.8$ \\[-1ex]
 NLO (pb) & $726.9$ & $746.4$ & $826.4$ & $811.4$ \\[-1ex]
 NNLO (pb)& $794.0$ & $808.0$ & $833.8$ & $822.4$ \\ \hline
\end{tabular}
\end{center}
\vspace*{-0.6cm}
\caption{
Total cross section at $\sqrt{s}=13$~TeV in the pole scheme with $M_t = 173.3\text{ GeV}$  and in the \ms scheme with $\mbar = 163.7\text{ GeV}$. 
Results obtained with different values of the auxiliary scales $\mu_R, \mu_F$ and
$\mu_m$.
}
\label{table:scales}
\end{table}
}

We note that the \ms scheme results with $\muR=\muF=\mbar$ and $\mum=\mbar/2$
have a reduced perturbative convergence (the $K$-factors are $K_\text{NLO}=1.53$ and
$K_\text{NNLO}=1.08$),
and, at each perturbative order, they are quantitatively very similar to the pole scheme results with $\muR=\muF=\Mt$. This similarity is not unexpected, since if 
$\mum=\mbar/2$ the value of the top-quark mass $m_t(\mum)$ that is used in the \ms
scheme computation is almost equal to \Mt.

By contrast, if the central scales $\muR = \muF = \Mt/2$ are used within the pole scheme (as suggested in Ref.~\cite{Czakon:2016dgf}), the computation shows a faster perturbative convergence (the $K$-factors are $K_\text{NLO}=1.31$ and
$K_\text{NNLO}=1.01$)
and, at each perturbative order, the results are quantitatively very similar to the 
\ms scheme results with $\muR=\muF=\mum=\mbar$. Within the pole scheme a faster perturbative convergence is achieved by lowering the value of \muR (and \muF, to a smaller extent), thus increasing the cross section results at lower orders.

We draw some overall conclusions from our comparison between pole and \ms scheme results for the total cross section up to NNLO. At each perturbative order the two schemes give results that are consistent within the corresponding scale uncertainties,
and, in particular, the results are quite similar at NNLO.
The pattern of slower or faster apparent perturbative convergence depends on both the mass renormalisation scheme and the actual values of the auxiliary scales. This is a consequence of the fact that at low orders the radiative corrections are relatively large and the results have a sizeable dependence on \muR 
and on \mum within the \ms scheme. Using the central scales of 
Table~\ref{table:totalXS}, the perturbative convergence of the \ms scheme is faster
than that of the pole scheme. An opposite (or intermediate) pattern of perturbative convergence can be obtained by using other central scales that are within the range of scale variations that is used in Table~\ref{table:totalXS}. Therefore, in the case of the total cross section, the scale setting that is used in Table~\ref{table:totalXS}
is sufficiently representative of the scale dependence effects that can affect the comparison between the pole and \ms schemes. We expect that this remains true for other \ttb production observables. Therefore, in our subsequent comparison of the pole and \ms schemes for single-differential cross sections we still use the scale setting of Table~\ref{table:totalXS}.

\subsection{Differential cross sections}
\label{sec:diff}

\begin{figure}
\begin{center}
\includegraphics[width=0.49\textwidth]{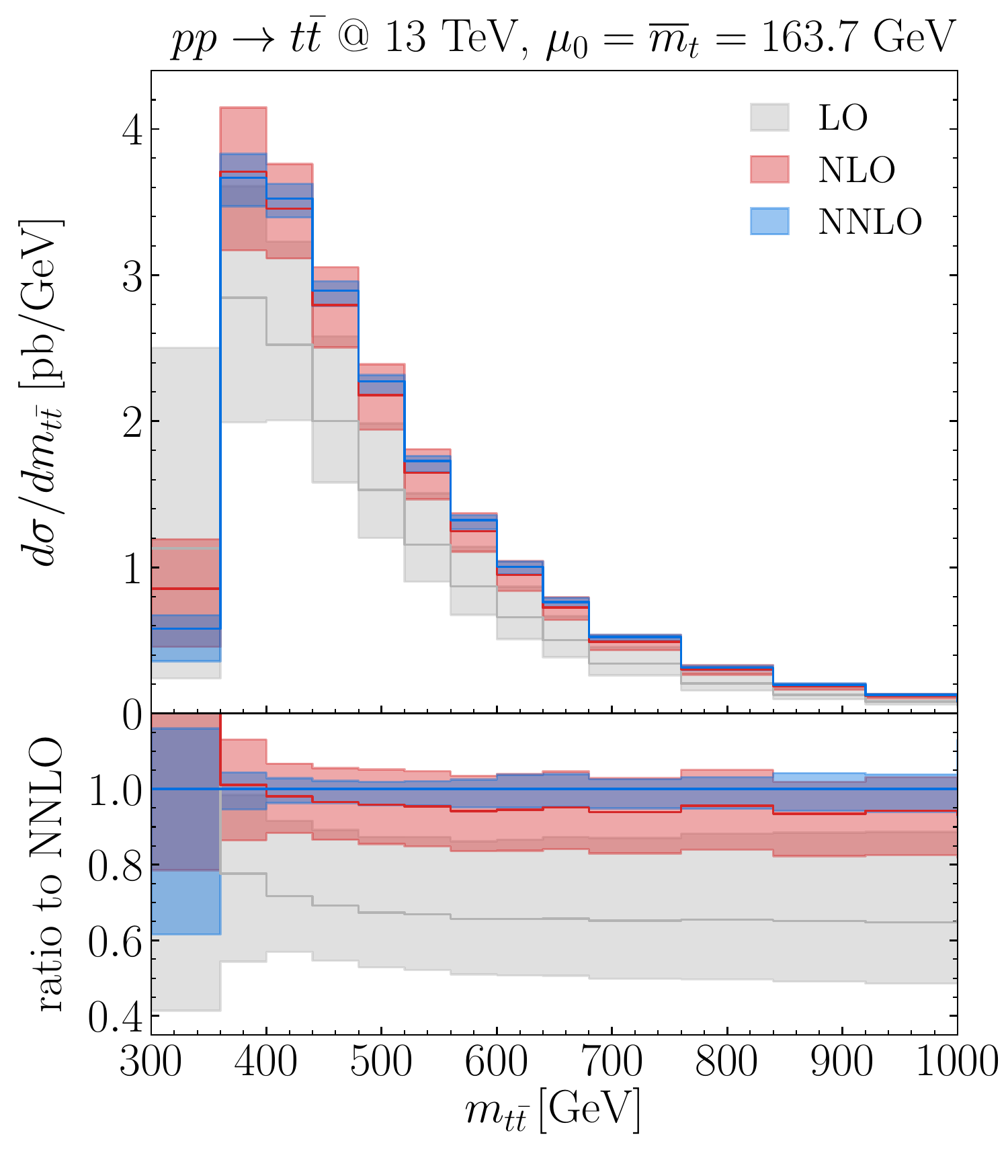}
\includegraphics[width=0.49\textwidth]{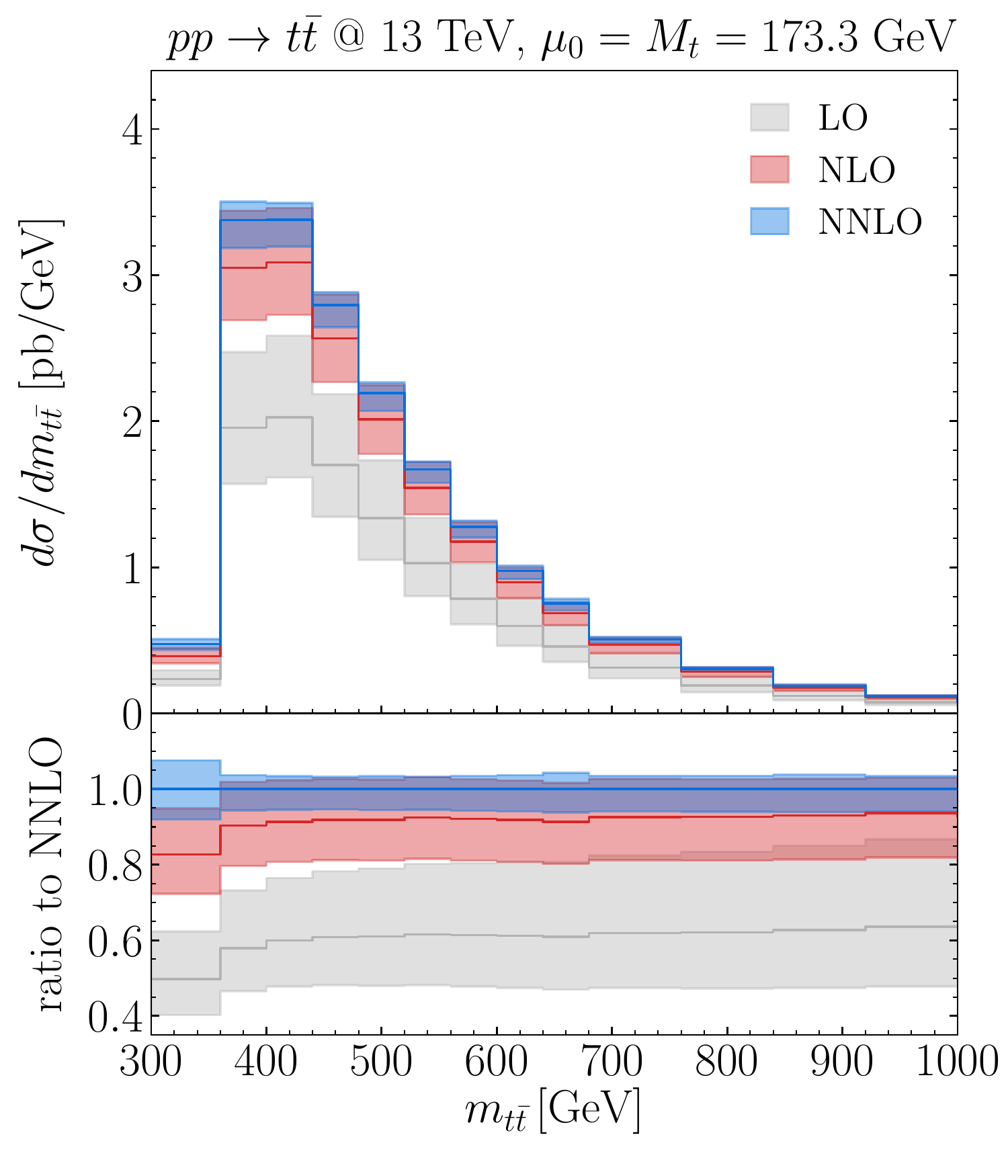}
\end{center}
\vspace{-5ex}
\caption{\label{fig:mtt}
Top-quark pair invariant-mass distribution at LO (gray), NLO (red) and NNLO (blue) within the \ms (left) and pole (right) schemes. The lower panel shows the ratio to the corresponding NNLO result. 
The values of the top-quark masses and of the auxiliary scales are the same as in Table~\ref{table:totalXS}.
}
\end{figure}

\begin{figure}
\begin{center}
\includegraphics[width=0.49\textwidth]{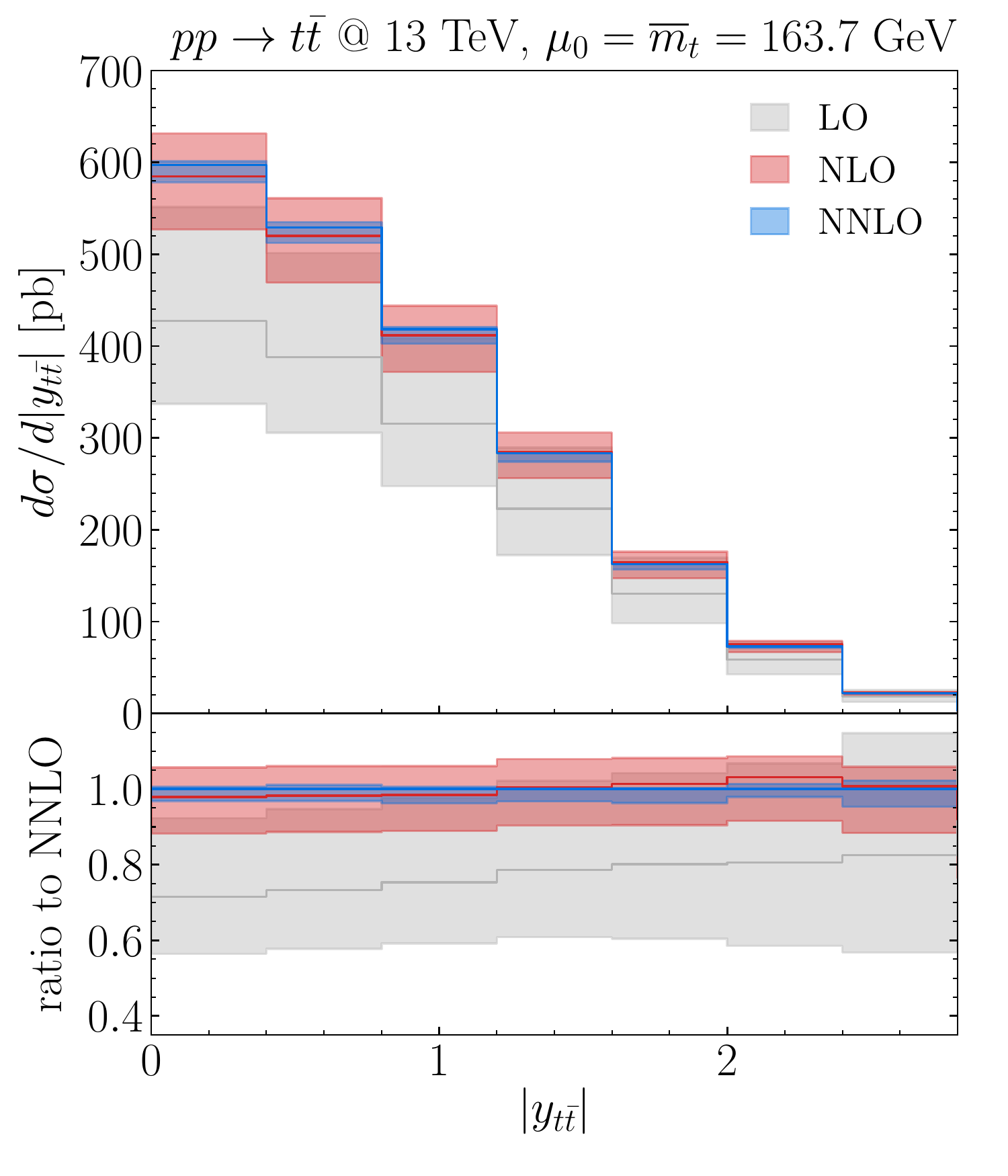}
\includegraphics[width=0.49\textwidth]{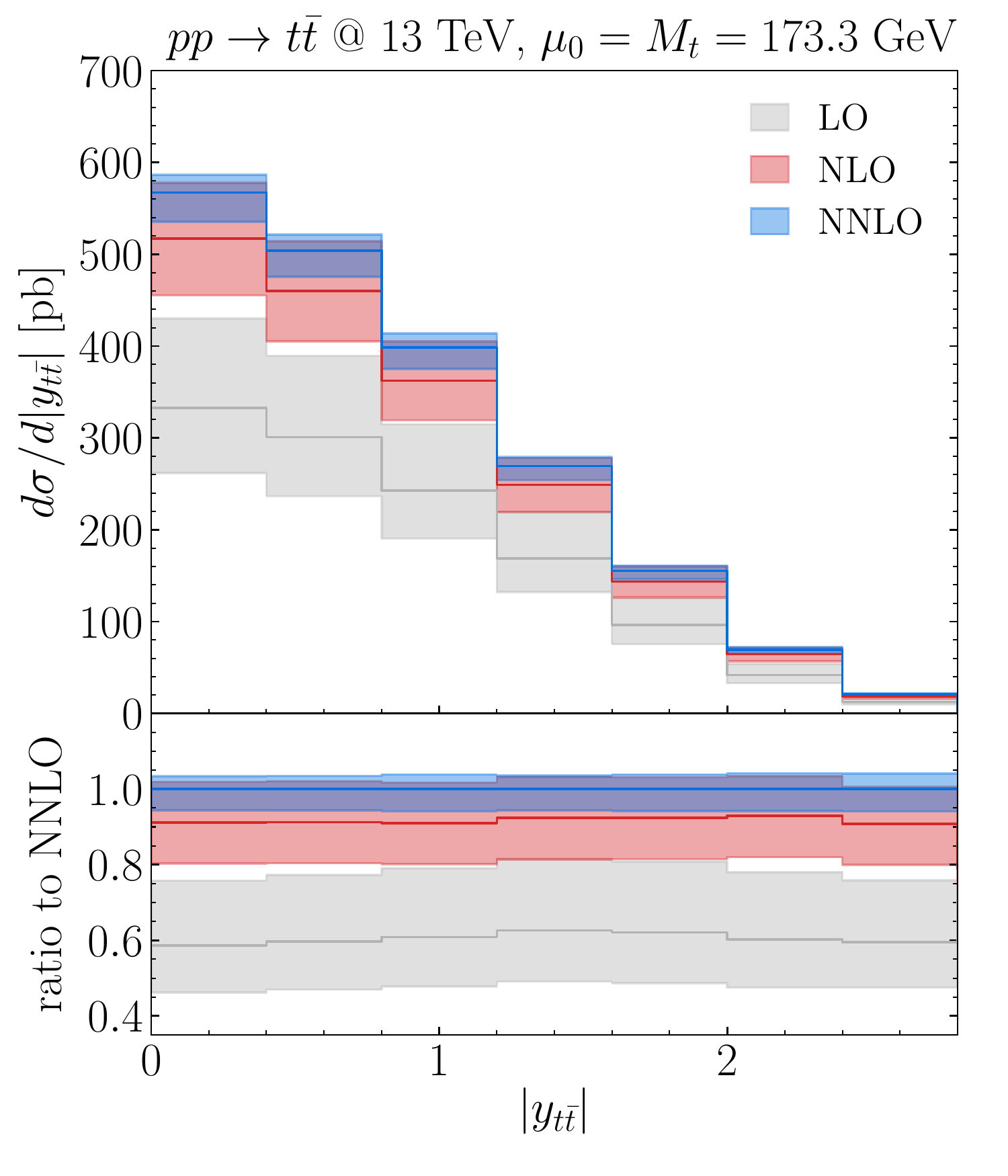}
\end{center}
\vspace{-5ex}
\caption{\label{fig:ytt}
Top-quark pair rapidity distribution at LO (gray), NLO (red) and NNLO (blue) within the \ms (left) and pole (right) schemes. The lower panel shows the ratio to the corresponding NNLO result.
The values of the top-quark masses and of the auxiliary scales are the same as in Table~\ref{table:totalXS}.
}
\end{figure}

\begin{figure}
\begin{center}
\includegraphics[width=0.49\textwidth]{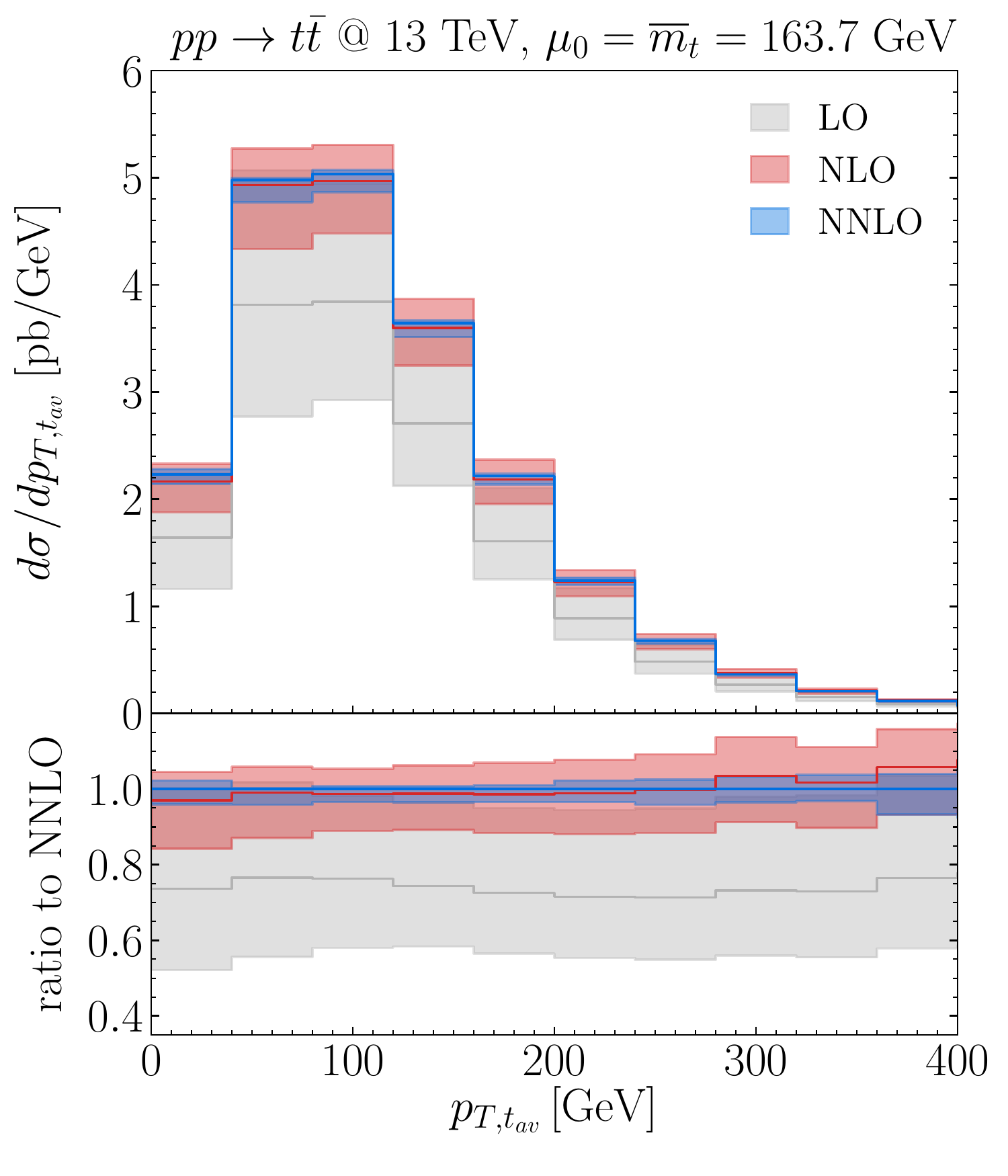}
\includegraphics[width=0.49\textwidth]{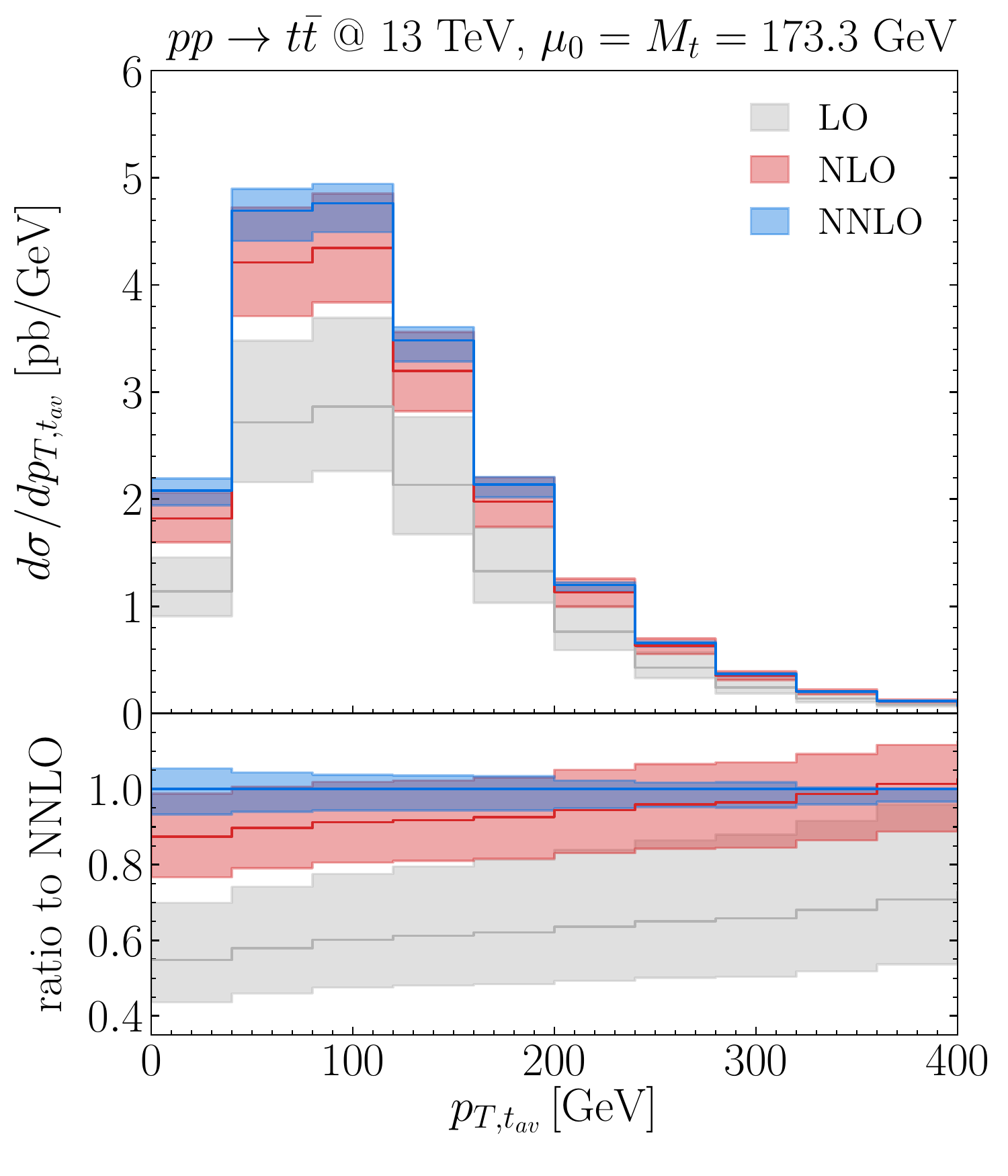}
\end{center}
\vspace{-5ex}
\caption{\label{fig:pTav}
Top-quark ($t$ and $\bar t$ average) transverse momentum distribution at LO (gray), NLO (red) and NNLO (blue) within the \ms (left) and pole (right) schemes. The lower panel shows the ratio to the corresponding NNLO result.
The values of the top-quark masses and of the auxiliary scales are the same as in Table~\ref{table:totalXS}.
}
\end{figure}

\begin{figure}
\begin{center}
\includegraphics[width=0.49\textwidth]{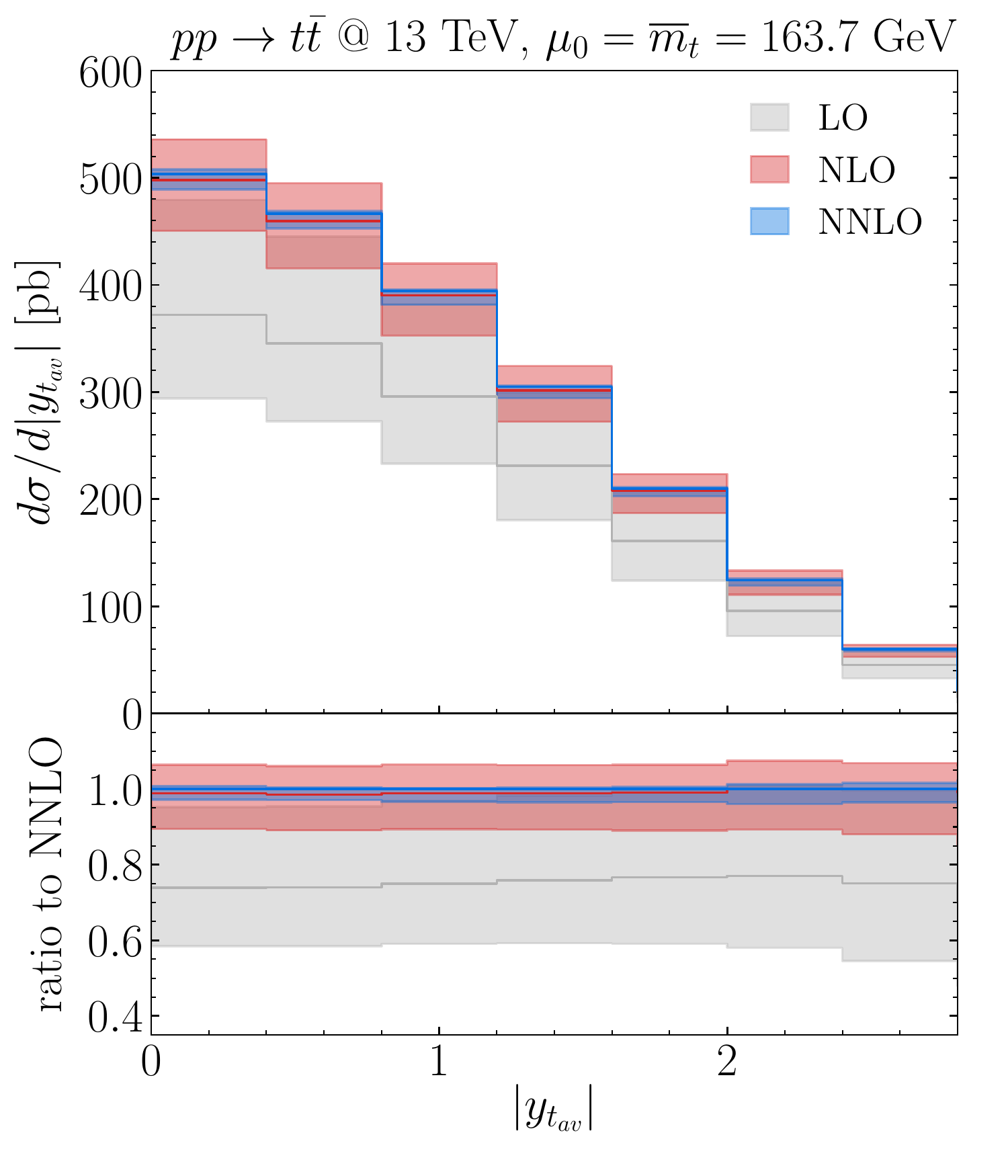}
\includegraphics[width=0.49\textwidth]{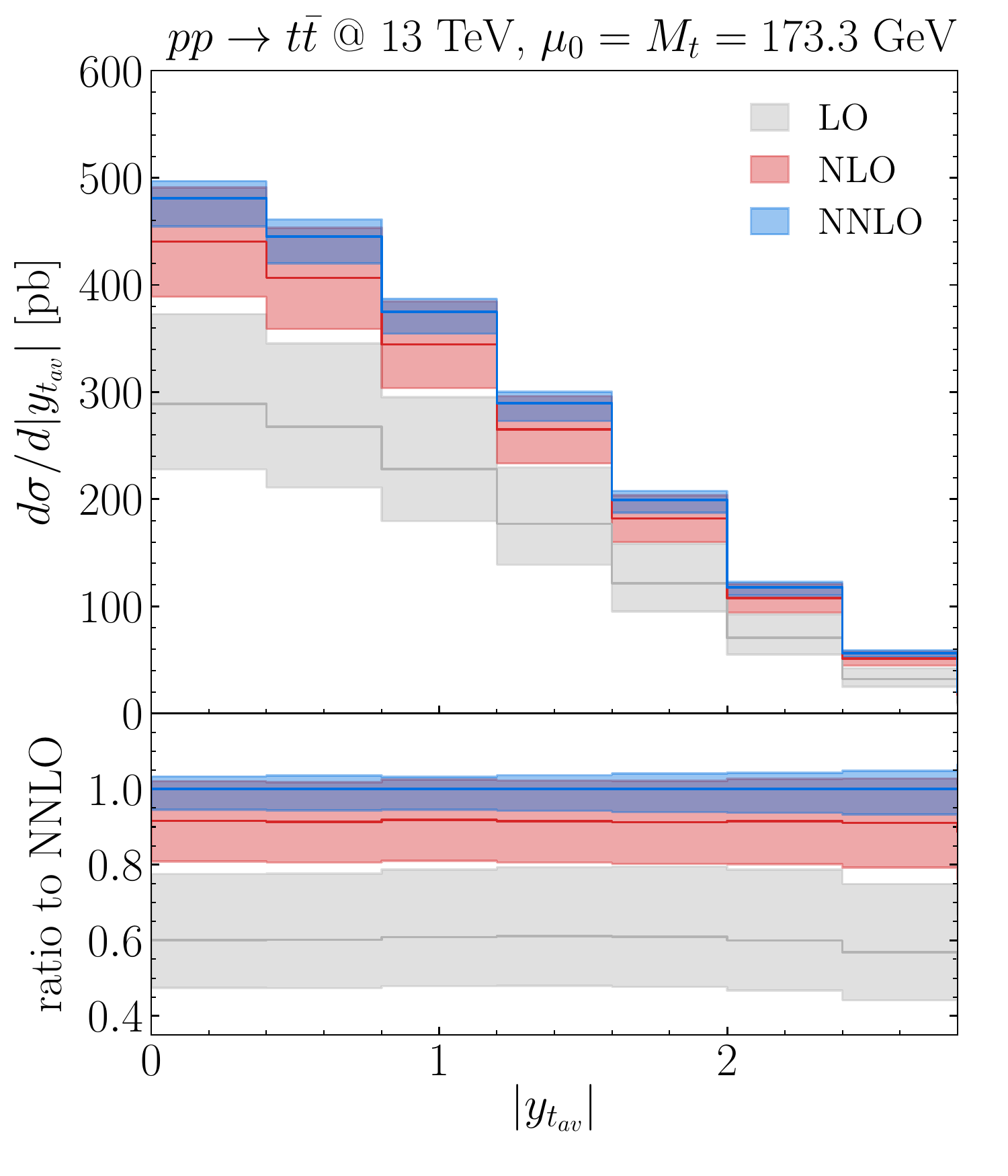}
\end{center}
\vspace{-5ex}
\caption{\label{fig:yav}
Top-quark ($t$ and $\bar t$ average) rapidity distribution at LO (gray), NLO (red) and NNLO (blue) within the \ms (left) and pole (right) schemes. The lower panel shows the ratio to the corresponding LO result.
The values of the top-quark masses and of the auxiliary scales are the same as in Table~\ref{table:totalXS}.
}
\end{figure}

We now move to consider differential results.
At the centre-of-mass energy ${\sqrt s}=13$~TeV,
we have computed the differential cross sections $d\sigma/dX$ with respect to the following variables $X$: the invariant mass
\mtt (\reffig{fig:mtt}) and the rapidity $y_{\ttb}$
(\reffig{fig:ytt}) of the \ttb pair, the average value of the transverse momentum $p_{T,t_\text{av}}$ (\reffig{fig:pTav}) and rapidity
$y_{t_\text{av}}$ (\reffig{fig:yav}) of the top quark and antiquark.
The perturbative results up to NNLO in the \ms and pole schemes are presented in the left and right panels, respectively. The values of the top-quark masses and of the auxiliary scales are the same as used in Table~\ref{table:totalXS}.
In particular, the scale uncertainty bands refer to the 15-point scale variations
for the \ms scheme and the 7-point scale variations for the pole scheme.
In the lower panels of Figs.~\ref{fig:mtt}--\ref{fig:yav} we show the ratios
of the perturbative results with respect to the central NNLO result in the corresponding mass renormalisation scheme.
The LO and NLO differential cross sections with respect to \mtt, $p_{T,t_\text{av}}$ and $y_{t_\text{av}}$ in the \ms scheme were computed in 
Ref.~\cite{Dowling:2013baa}, and our LO and NLO results are consistent with them. 


Comparing the \ms and pole scheme results in Figs.~\ref{fig:mtt}--\ref{fig:yav},
we observe some overall features that are fully analogous to those observed by considering the $t {\bar t}$ total cross section.
The \ms scheme results typically have smaller $K$-factors and a larger overlap between the scale variation bands at subsequent perturbative orders. At NLO, these features were already noticed in Ref.~\cite{Dowling:2013baa}. Going to NNLO we note, in particular, that within the \ms scheme the NNLO scale uncertainty band 
is contained within the NLO band for most of the kinematical regions considered in
Figs.~\ref{fig:mtt}--\ref{fig:yav}. These overall comments apply to the results in 
Figs.~\ref{fig:mtt}--\ref{fig:yav} with the exception of the \mtt distribution
at low values of \mtt. The low-\mtt region deserves further specific comments.

The invariant-mass distribution for on-shell $t {\bar t}$ production 
has a physical threshold at the value $\mtt=2 \Mt$, and it has a sharply
increasing behaviour just above the threshold.
Within the pole scheme (Fig.~\ref{fig:mtt} right), this behaviour is fulfilled order-by-order in perturbation theory. As we have already mentioned in Section~\ref{sec:msbar},
this physical behaviour is spoiled by the fixed-order perturbative expansion in the \ms scheme. This is the consequence of the identification of the top-quark mass with the 
\ms mass $m_t(\mum)$ in the computation of the perturbative on-shell cross sections
$\sigma^{(i)}$ of Eqs.~(\ref{eq:barlo})--(\ref{eq:barnnlo}). The order-by-order computation of the \mtt distribution in the \ms scheme produces a threshold at the value $\mtt=2 m_t(\mum)$, and such threshold is unphysical for a twofold reason: the threshold value differs from $2 \Mt$, and it depends on the unphysical
(arbitrary) auxiliary scale \mum. Besides producing an unphysical threshold,
the \ms scheme perturbative expansion 
also produces instabilities at subsequent perturbative orders since the mass derivatives $(m \partial_m)^k \sigma^{(i)}$ in Eqs.~(\ref{eq:barnlo}) and 
(\ref{eq:barnnlo}) are very large close to the threshold region. These perturbative
instabilities are clearly visible in Fig.~\ref{fig:mtt} by comparing the \ms 
and pole scheme results at low values of \mtt: in the first and, partly, second \mtt bins the \ms scheme results have larger $K$-factors (at both NLO and NNLO) and wider scale uncertainty bands.
At NLO this unstable behaviour has also been observed and discussed in Ref.~\cite{Dowling:2013baa}.

Owing to these features, the formal replacement of the pole mass with 
the \ms mass
is not a justified (and, thus, not a recommended) procedure for the 
perturbative computation of the invariant-mass distribution close to 
its on-shell production threshold.
However, we note that
the unphysical perturbative behaviour of the \ms scheme computation at 
low values of $m_{\ttb}$
can be partly alleviated by considering three related (and correlated) aspects: the use of large bin sizes, the inclusion of variations of the renormalisation scale \mum, the computation of higher-order contributions. We comment on these aspects in turn.

We recall that, in our \ms scheme computation of the \mtt cross section, the \ms mass $m_t(\mum)$  varies (due to \mum variations) in the range 
$155.5\text{~GeV}\ltap m_t(\mum) \ltap 173.3\text{~GeV}$
(we also recall that the upper value coincides with the value of \Mt).
The first \mtt bin
in Fig.~\ref{fig:mtt} extends from 300~GeV to 360~GeV, so that the unphysical thresholds are always included in the first bin. The reduction of the bin size will amplify the unphysical behaviour of the \ms scheme computation.

At low values of \mtt the scale dependence of the \ms scheme results is very large, thus leading to a sizeable perturbative uncertainty and, therefore, mitigating the effect of the unphysical features at the quantitative level. We note that at low values of \mtt this scale dependence is largely dominated by the effect of
\mum variations (which, through
Eq.~(\ref{eq:xs_MSbar}), changes the position of the \mtt threshold), while
the variations of \muR and \muF lead to much smaller quantitative effects.
This is in contrast with the \ms scheme results in the region of higher values of 
\mtt, where (analogously to the total cross section results in the second and third columns of Table~\ref{table:totalXS}) the 15-point and 7-point (i.e., by fixing 
$\mum = \mbar$) scale variations produce quantitatively similar scale uncertainties.

\begin{figure}
\begin{center}
\includegraphics[width=0.48\textwidth]{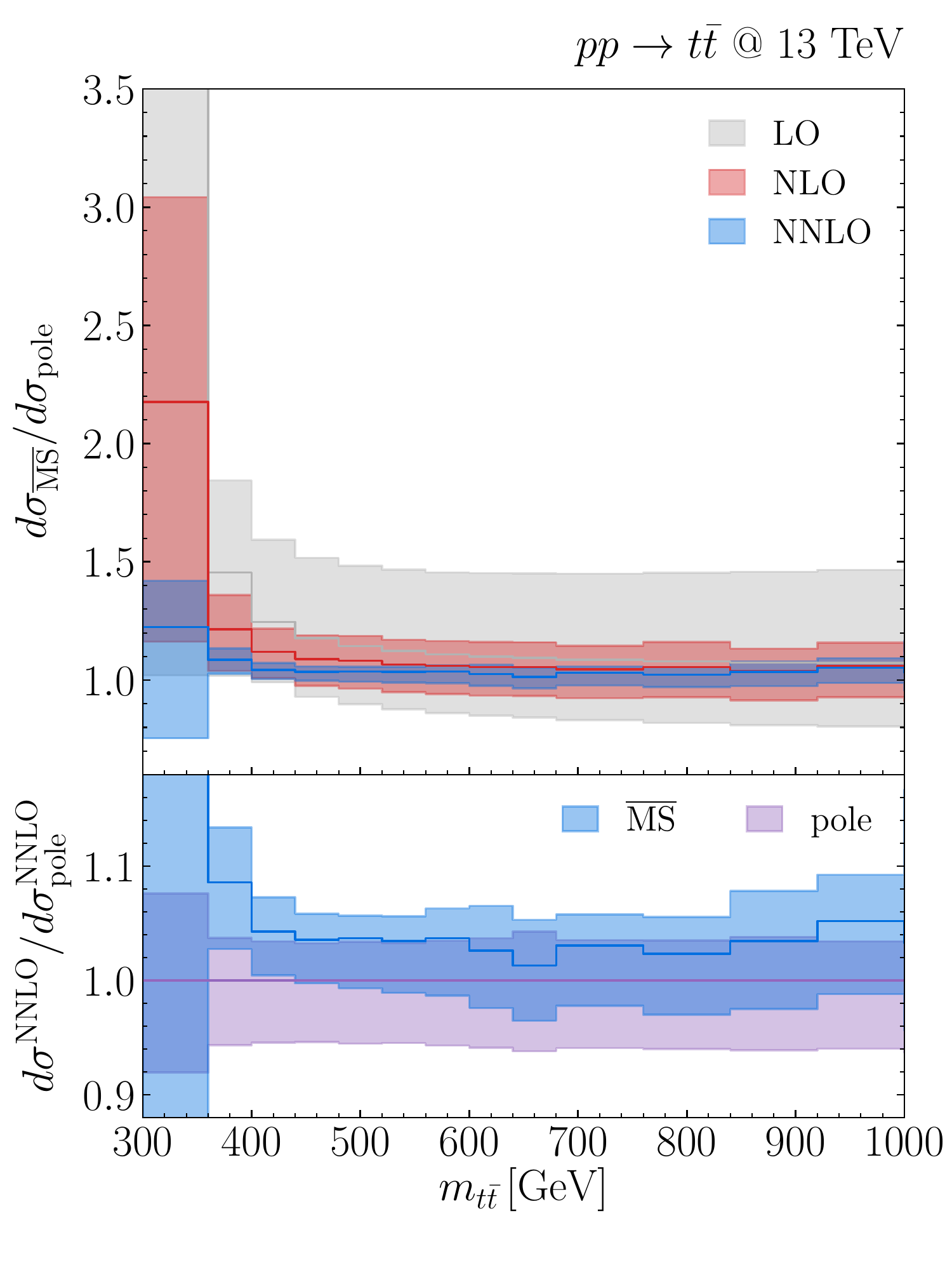}
\includegraphics[width=0.48\textwidth]{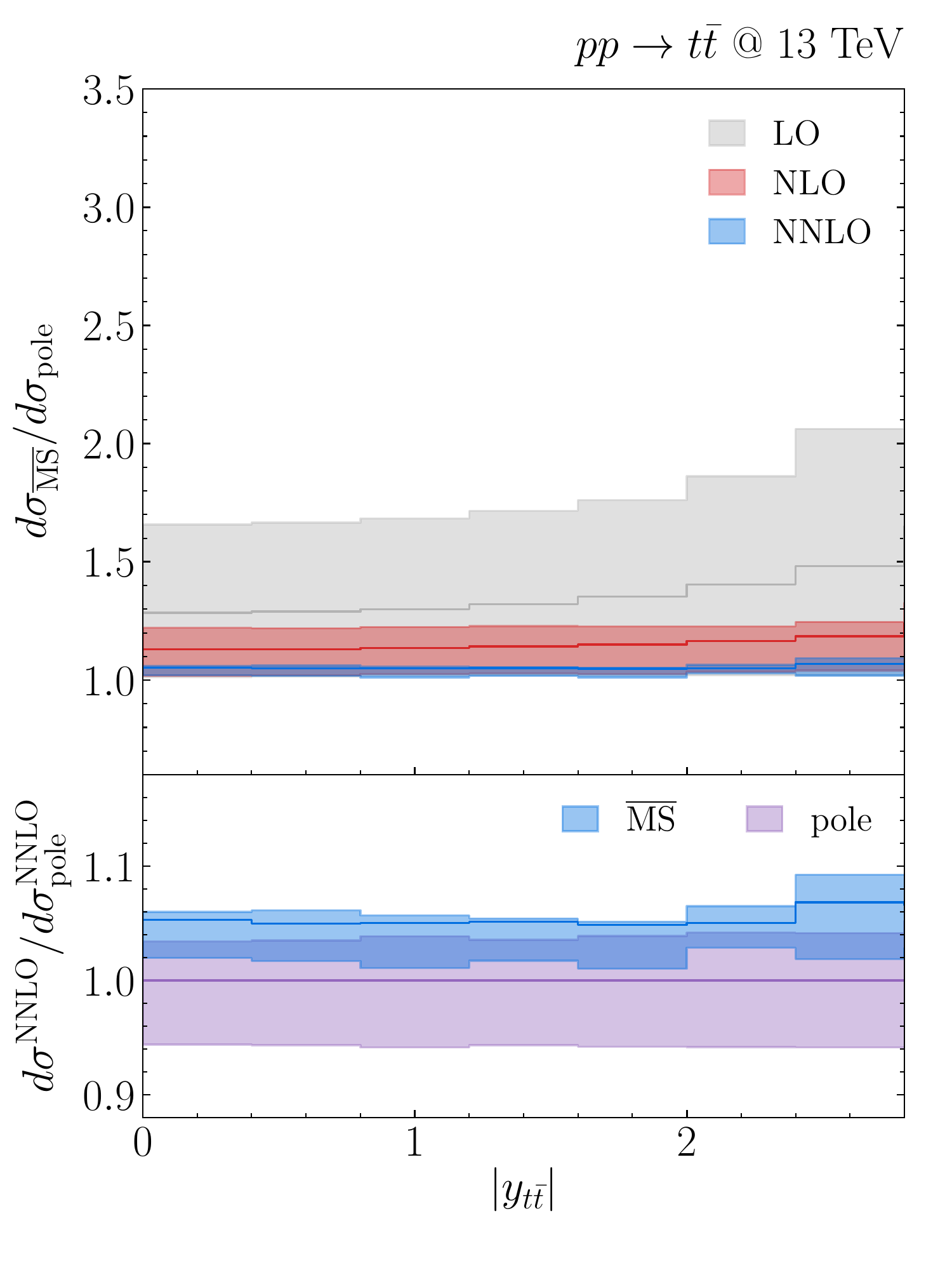} \\
\vspace*{-0.5cm}
\includegraphics[width=0.48\textwidth]{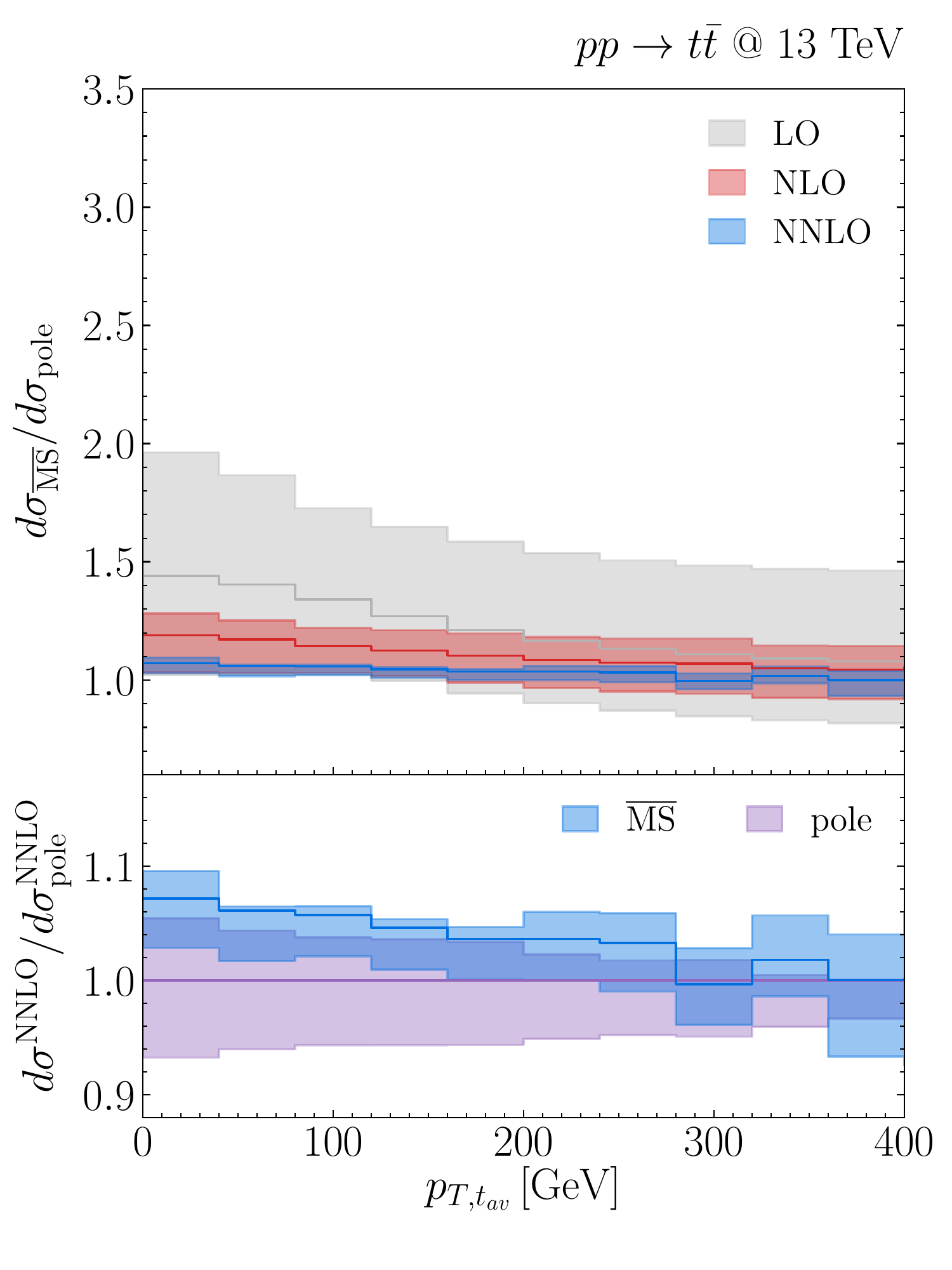}
\includegraphics[width=0.48\textwidth]{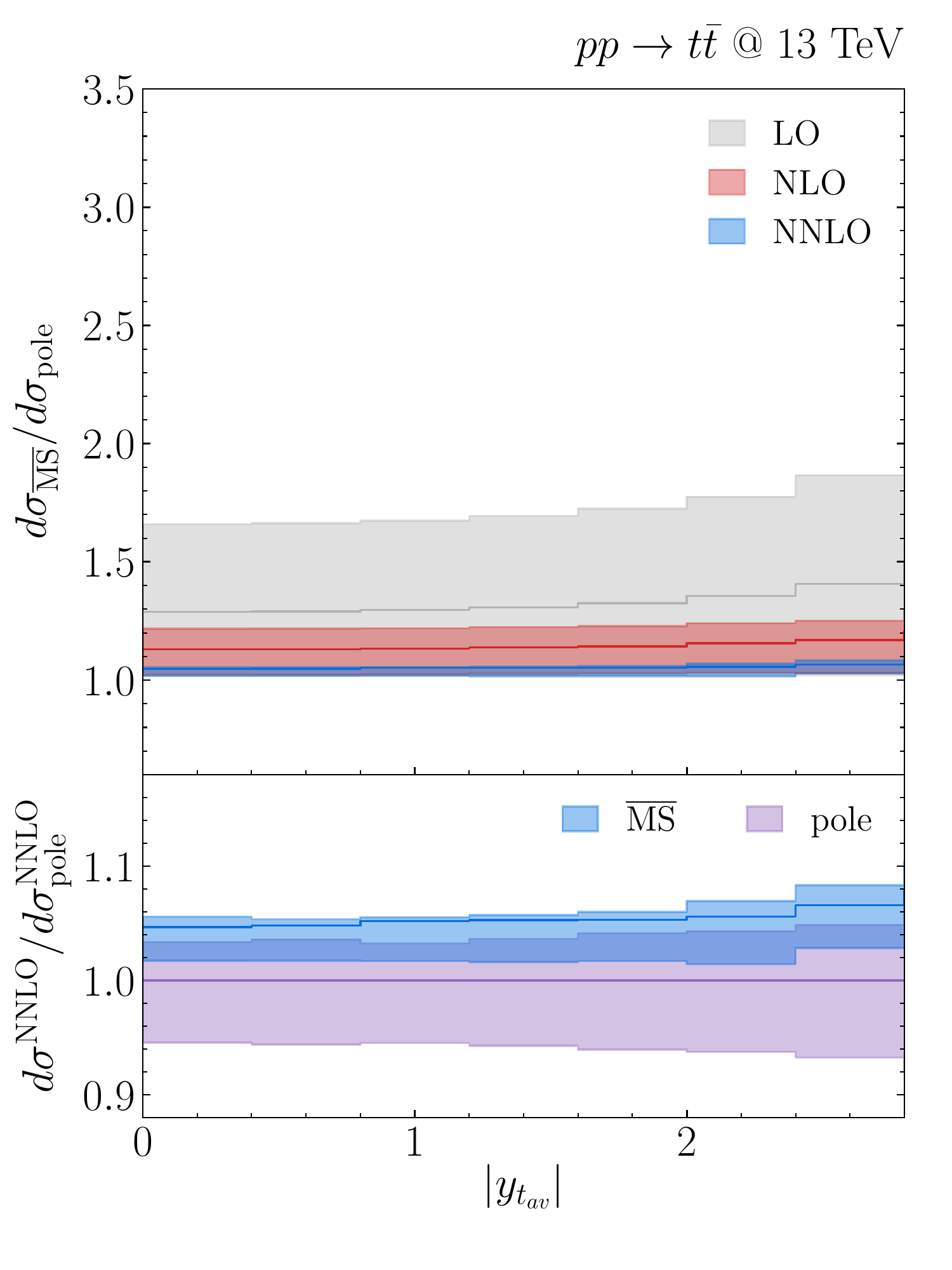}
\end{center}
\vspace{-6ex}
\caption{\label{fig:ratios}
Ratio between the \ms and pole scheme predictions at LO (gray), NLO (red) and NNLO (blue) for the differential distributions in 
Figs.\ref{fig:mtt}--\ref{fig:yav}.
The numerator is the \ms scheme result with its 15-point scale variation,
while the denominator is the central result (at each corresponding order) in the pole scheme. 
The lower panels show only the NNLO ratio, including the uncertainty band of the pole scheme result (purple).
}
\end{figure}

The unphysical features of the \ms scheme computation at low \mtt are due to the low-order perturbative expansion in Eqs.~(\ref{eq:barlo})--(\ref{eq:barnnlo}). Owing to the formal all-order equality in Eq.~(\ref{eq:all}), these unphysical features tend to `disappear' by expanding the \ms scheme cross section $\bar \sigma$ at a `sufficiently' high order (see the results in Fig.~\ref{fig:ratios} and related accompanying comments).

Considering the behaviour of the \mtt distribution near the threshold region, we note that the perturbative computation
in the pole scheme also leads to enhanced radiative corrections, which are of dynamical origin.
All-order resummed calculations of Coulomb-type radiative corrections combined with effects of the finite width
$\Gamma_t$ of the top quark were presented in Refs.~\cite{Hagiwara:2008df,Kiyo:2008bv,Ju:2019mqc,Ju:2020otc}.
In particular, the calculation of Refs.~\cite{Ju:2019mqc,Ju:2020otc} leads to an increase of about 9\% of the NNLO differential cross section integrated over the bin where $300~\text{GeV} <  \mtt < 380~\text{GeV}$
(larger resummation effects occur for the detailed shape of $d\sigma/d \mtt$ over a more restricted region of size 
$\Delta \mtt \sim \Gamma_t$ around the on-shell threshold at $\mtt=2\Mt$). We remark on the fact that the dynamical effects
considered in Refs.~\cite{Hagiwara:2008df,Kiyo:2008bv,Ju:2019mqc,Ju:2020otc} are unrelated to those produced by the change of mass renormalisation schemes
from the pole to the \ms scheme.

We note that our comments and discussion on the unphysical features of the 
invariant-mass distribution at low values of \mtt similarly apply to other differential distributions in kinematical regions that are sensitive to thresholds related to on-shell $t {\bar t}$ production. For instance, this is the case for the 
$y_{\ttb}$ and $y_{t_\text{av}}$ differential cross sections in the very high rapidity region (specifically, at values of $|y_{\ttb}|$ and $|y_{t_\text{av}}|$
that are larger than those considered in the results of Figs.~\ref{fig:ytt} and
\ref{fig:yav}).

To perform a more direct comparison of the \ms and pole scheme results 
for the differential cross sections in Figs.~\ref{fig:mtt}--\ref{fig:yav}, we compute the ratio between them at each perturbative order for each of the distributions.
The ratios are presented in Fig.~\ref{fig:ratios}, where the lower panels show in more detail the NNLO results and their uncertainty bands in both the \ms and pole schemes.
We see that for all the distributions the results in the two schemes are consistent at LO, NLO and NNLO within the corresponding scale uncertainties. In particular, and in relation to our previous discussion of the low-\mtt region, we note that the quantitative effect of \mum variations in the \ms scheme is particularly relevant to get consistency with the pole scheme results for the invariant-mass distribution.
At NNLO the scale uncertainty band of the \ms scheme results is typically of the same size (see the \mtt and $p_{T,t_\text{av}}$ cross sections) or smaller 
(see the $y_{\ttb}$ and $y_{t_\text{av}}$ cross sections) than the corresponding band in the pole scheme.
In the cases of the $y_{\ttb}$ and $y_{t_\text{av}}$ distributions, we also see that the NNLO scale band in the \ms scheme is highly asymmetric with respect to its central value (which, therefore, is evaluated close to a region of local minimal sensitivity
to scale variations), consistently with the similar behaviour of the total cross section in Table~\ref{table:totalXS}. 

In Fig.~\ref{fig:ratios} we also observe that the shape differences between the \ms
and pole schemes are significantly reduced by the inclusion of high-order corrections,
and they are already quite small at NNLO. Moreover, and importantly, 
in all the kinematical regions of Fig.~\ref{fig:ratios}
we note a sizeable overlap between the \ms and pole scheme uncertainty bands at NNLO:
this fact shows the expected similarity
between the two schemes once enough perturbative orders are included in the calculation.

\subsection{Comparison with CMS data and running-mass effects}
\label{sec:running}

Up to now we have
presented perturbative calculations in the \ms scheme by using values of the renormalisation scale 
\mum that are of the order of the top-quark mass. In the following we refer to these calculations as predictions with a
\textit{fixed} \ms mass, since the scale \mum is not necessarily related to the characteristic scale of the differential cross section under consideration. In the remaining part of this Section we also consider QCD predictions that use a \textit{running} \ms mass, namely, perturbative calculations in which the \ms mass $m_t(\mum)$ is evaluated at a dynamical value of  \mum that is related to the hard-scattering scale of the differential cross section. Specifically, we consider QCD predictions for the invariant-mass cross section
$d\sigma/d \mtt$, since its characteristic scale is \mtt, which can be parametrically much larger than the top-quark mass.

In Ref.~\cite{Sirunyan:2019jyn} the CMS Collaboration performed a measurement of the invariant-mass distribution for \ttb
production based on $35.9~\text{fb}^{-1}$ of LHC data at the centre-of-mass energy ${\sqrt s}=13$~TeV.
The measurement was then compared with QCD predictions in the \ms scheme to the purpose of performing a determination of the top-quark mass.

The procedure used in Ref.~\cite{Sirunyan:2019jyn} by the CMS Collaboration is as follows. The theoretical results for 
$d\sigma/d \mtt$ are obtained by using the NLO QCD calculation~\cite{Dowling:2013baa}
in the \ms scheme with a fixed  scale $\mum=\mbar$, and treating $m_t(\mbar)=\mbar$ as a free parameter.
The value of $\mbar$ in each invariant-mass bin is then determined by comparing these theoretical predictions with the
data point in the same bin. The fitted value $\mbar^{(k)}$ of $\mbar$ in the $k^{\text{th}}$ bin is then used to compute 
$m_t(\mu_k)$ at the characteristic invariant-mass scale $\mu_k$~\cite{Sirunyan:2019jyn} of the corresponding bin.
The computation of $m_t(\mu_k)$ from $m_t(\mbar^{(k)})=\mbar^{(k)}$ is performed by using the evolution equation (\ref{eq:eveq})
at LO. The final result of the CMS Collaboration~\cite{Sirunyan:2019jyn} is that
the $\mu_k$ dependence of the determined values of $m_t(\mu_k)$ agrees (within theoretical and experimental errors)
with the expectation from the evolution equation (\ref{eq:eveq}) at LO.

The final result of Ref.~\cite{Sirunyan:2019jyn} implies that the fitted values of $\mbar^{(k)}$ in the various invariant-mass bins are consistent (within errors) with a single common (i.e., bin-independent) value. In view of this, we conclude that the CMS data of
Ref.~\cite{Sirunyan:2019jyn} on $d\sigma/d \mtt$ are consistent with the NLO QCD predictions in the \ms scheme as obtained by using a fixed value $m_t(\mbar)=\mbar$ of the \ms mass, namely, without introducing dynamical effects due to the running
of the \ms mass.\footnote{Indeed, a fixed value of $\mum=\mbar$ is used in the NLO QCD calculation of  Ref.~\cite{Sirunyan:2019jyn}
  for all the invariant-mass bins.}
Therefore, the analysis performed in Ref.~\cite{Sirunyan:2019jyn} has no direct sensitivity to running-mass effects,
contrary to what is stated therein.

In Section~\ref{sec:total} we have computed the $t{\bar t}$ total cross section by fixing $\mu_m=\mbar$
and using the corresponding fixed value, $m_t(\mbar)=\mbar$, of the \ms mass.
Such QCD predictions can be used to determine the value of $\mbar$ through a comparison
with data for the $t{\bar t}$ total cross section (see, e.g., Ref.~\cite{Alekhin:2018pai}),
but such comparison cannot be used to measure the running of the \ms mass of the top quark.
Analogously, in the case of differential cross sections, QCD predictions in the \ms scheme that are obtained
by using a fixed value of the mass renormalization scale $\mu_m$ and, hence, a fixed value of the \ms mass
(such as the NLO calculation used in Ref.~\cite{Sirunyan:2019jyn}) can be exploited to determine
this value, but they cannot be exploited to study the scale dependence and the running
of the \ms mass. The investigation of this behaviour requires (at least) the use of QCD calculations with a running
(i.e., not fixed at a unique value) renormalization scale $\mu_m$.

\begin{figure}[t]
\begin{center}
\includegraphics[width=0.49\textwidth]{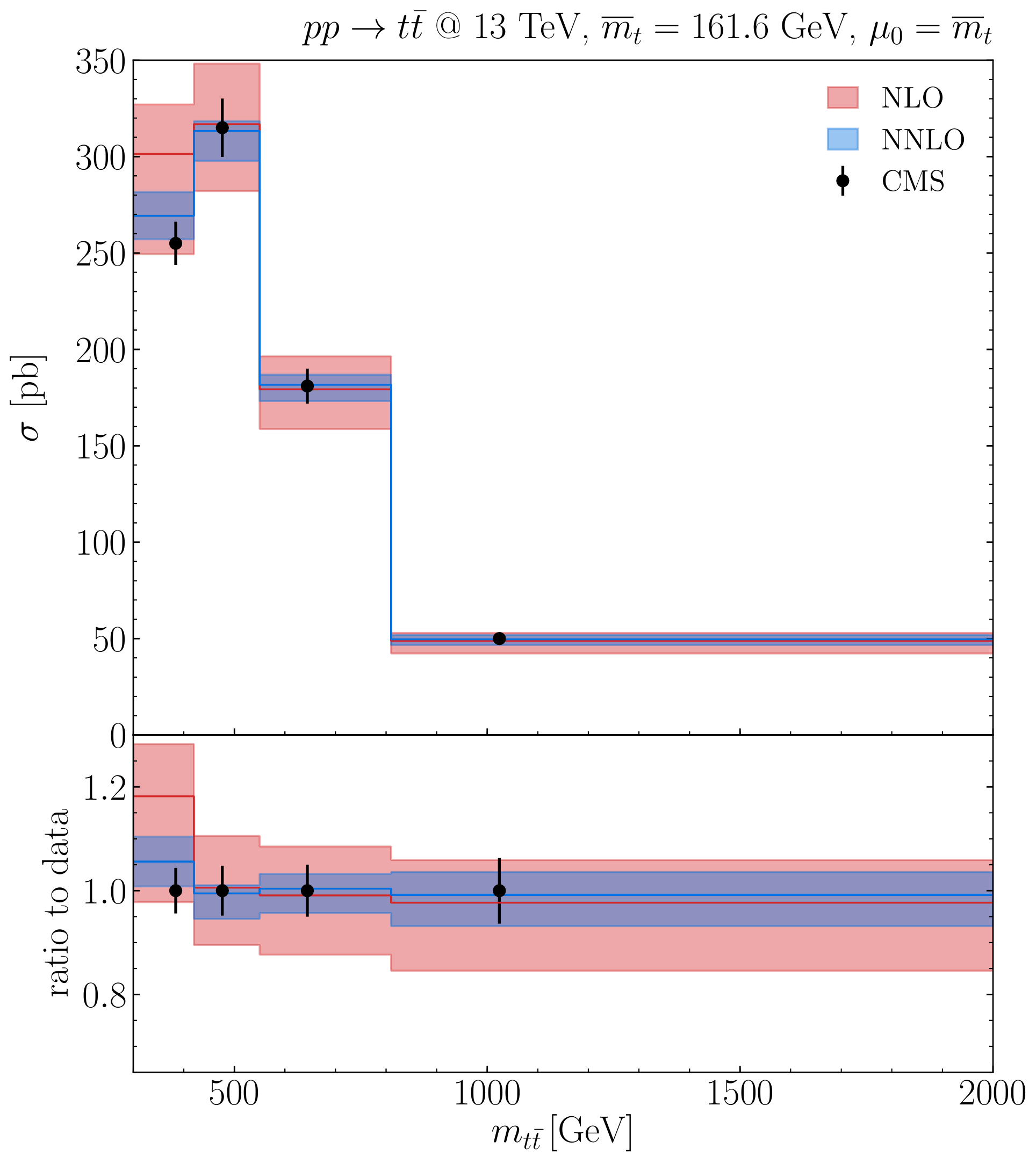}
\includegraphics[width=0.49\textwidth]{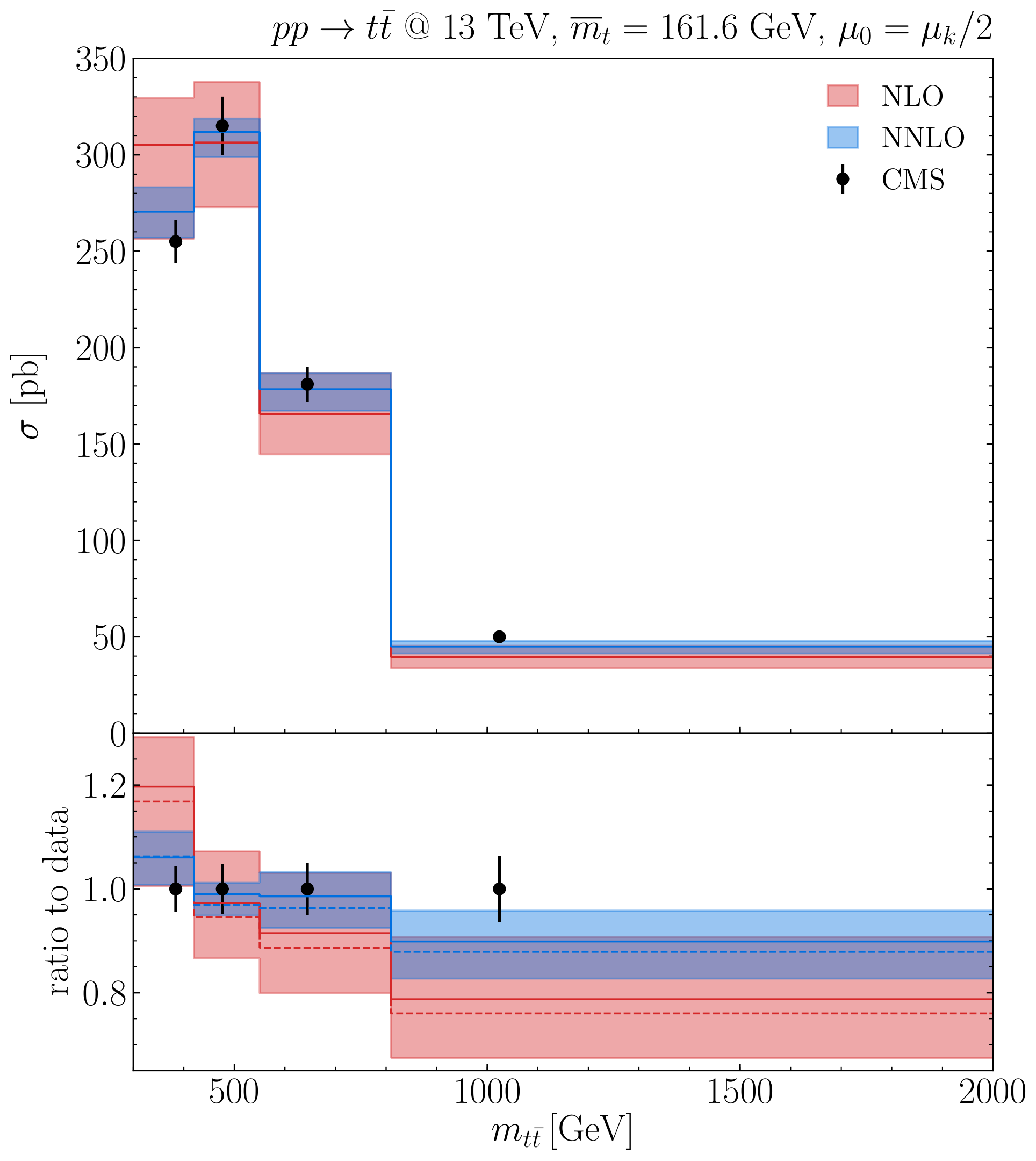}
\end{center}
\vspace{-4ex}
\caption{\label{fig:mtt-CMS-mt}
The invariant-mass distribution of the top-quark pair at NLO (red) and NNLO (blue) in the \ms scheme, and the result of the CMS measurement in Ref.~\cite{Sirunyan:2019jyn}. 
The theory uncertainty bands are obtained by performing 15-point scale variations around the central value $\mu_0$ of the auxiliary scales
$\muR, \muF$ and \mum.
In the left panel $\mu_0$ is fixed to $\mbar$, whereas in the right panel $\mu_0$ is dynamically set to $\mu_k/2$
($\mu_k$ is the centre of gravity of the cross section in the $k^{\text{th}}$ bin of \mtt).
In the lower right panel the dashed lines indicate the NLO and NNLO results obtained with $\mum = \mbar$ and $\muR = \muF = \mu_k/2$.
}
\end{figure}

Taking this fact into account, in the following we present a comparison (see Fig.~\ref{fig:mtt-CMS-mt})  of the CMS data of Ref.~\cite{Sirunyan:2019jyn}
with QCD predictions in the \ms scheme up to NNLO. The QCD predictions, which refer to the same binning as in the CMS measurement,
are obtained by using either a fixed (Fig.~\ref{fig:mtt-CMS-mt} left) or a running (Fig.~\ref{fig:mtt-CMS-mt} right) \ms mass,
as specified below. For both kinds of predictions,
we exactly follow the setup employed in Ref.~\cite{Sirunyan:2019jyn}: we use the ABMP16 PDF sets~\cite{Alekhin:2017kpj,Alekhin:2018pai} with $n_f=5$ massless-quark flavours, and the corresponding values of the QCD coupling, $\as(m_Z)=0.1191$ and $\as(m_Z)=0.1147$ at NLO and NNLO, respectively. 
The value of $\mbar$ is set to $161.6$~GeV, which corresponds to the result obtained by the CMS Collaboration~\cite{Sirunyan:2018goh} from a fit of the \ttb total cross section (using the same data set as in Ref.~\cite{Sirunyan:2019jyn})
based on NNLO predictions in the \ms scheme computed with the ABMP16 PDFs and the corresponding $\as$. We note that such value of
$\mbar$  is lower than the one
used to obtain all our previous results in the \ms scheme (e.g., the results in Figs.~\ref{fig:mtt} and \ref{fig:ratios}).
We also note that the value $\mbar = 161.6\text{ GeV}$ corresponds to the pole mass 
$\Mt=170.8\text{ GeV}$, by using the relation in Eq.~(\ref{eq:polemsbar}) at three-loop order.

The QCD predictions with a fixed \ms mass (Fig.~\ref{fig:mtt-CMS-mt} left) are computed analogously to those in Fig.~\ref{fig:mtt}.
We use the central value $\mu_0=\mbar$ for the three auxiliary scales \muR, \muF and \mum, and we consider the 15-point scale variations around this central value. At NLO this calculation corresponds to the one performed in Ref.~\cite{Sirunyan:2019jyn},
with the main difference that we include the uncertainties due to the variation of \mum by a factor of 2 around $\mu_0$
(\mum is kept fixed  to $\mbar$ in Ref.~\cite{Sirunyan:2019jyn}, though the effect of PDF uncertainties is considered therein).

The QCD predictions with a running \ms mass (Fig.~\ref{fig:mtt-CMS-mt} right) are computed by performing the 15-point scale variations 
around values of the central scale $\mu_0$ (for the three auxiliary scales \muR, \muF and \mum) of the order of 
$\mtt/2$, which is the characteristic hard-scattering scale of the differential cross section $d\sigma/d \mtt$. 
Specifically, in the $k^{\text{th}}$ invariant-mass bin we set  
$\mu_0 = \mu_k/2$
(setting directly 
$\mu_0=\mtt/2$
in our \ms scheme calculation is  more challenging from a 
computational point of view),
where $\mu_k$ is the centre of gravity of the \mtt cross section in the $k^{\text{th}}$ bin 
as computed by the CMS Collaboration~\cite{Sirunyan:2019jyn}.
The values of $\mu_k$ range from $\mu_1=384$~GeV in the $1^{\text{st}}$ bin to $\mu_4=1020$~GeV in the $4^{\text{th}}$ bin (see Table~1 in Ref.~\cite{Sirunyan:2019jyn}), and the corresponding values of the running-mass range from $m_t(\mu_1/2)=159.5$~GeV to $m_t(\mu_4/2)=149.0$~GeV (we use the evolution equation (\ref{eq:eveq}) at NNLO, as implemented in the package \CRunDec~\cite{Schmidt:2012az}).
We note that the fixed ($\mu_0=\mbar$) and dynamic ($\mu_0=\mu_k/2$) scales substantially differ only
in the high-$\mtt$ region (for instance, in the first bin $\mu_1/2=192$~GeV and $\mbar$ differ by less than
a factor of two, and both values are thus included within the scale variation range that we consider).
Therefore, our comparison between fixed-mass and running-mass predictions has the purpose
of investigating differences only at relatively high values of the invariant mass.\footnote{As discussed in Sect.~\ref{sec:diff}
    (see Fig.~\ref{fig:mtt} and accompanying comments),
    the use of the pole mass is preferred with respect to the \ms mass in
    the low-$m_{\ttb}$ region (e.g., in the first invariant-mass bin of
    Fig.~\ref{fig:mtt-CMS-mt}, where $m_{\ttb}<420$~GeV).}

From the theory--data comparison in Fig.~\ref{fig:mtt-CMS-mt} we see that 
the NNLO results in the \ms scheme with fixed or running masses are both in excellent agreement with data.
We also note that the agreement improves (especially for the predictions with the running mass at high values of \mtt)
in going from NLO to NNLO. As observed in Fig.~\ref{fig:ratios}, the NNLO predictions for $d\sigma/d \mtt$
in the pole and \ms schemes are consistent within their scale uncertainties. Therefore, we also note that the
level of agreement between data and NNLO theory is not a peculiarity of the results in the \ms scheme.
The CMS data are also fully compatible with the NNLO predictions in the pole scheme, provided they are obtained by using a value of pole mass 
\Mt that is consistent (according to Eq.~(\ref{eq:polemsbar}) at NNLO) with the value of $\mbar$ of the corresponding predictions in the \ms scheme.

Inspecting the results reported in Fig.~\ref{fig:mtt-CMS-mt}, we see that NNLO (and also NLO) \ms scheme predictions with a fixed and a running mass are consistent with each other within their scale uncertainties (in particular the two NNLO scale variation bands have a substantial overlap). In the highest invariant-mass bin of Fig.~\ref{fig:mtt-CMS-mt}, the data point agrees
better with the central NNLO prediction with a fixed \ms mass, which leads to a larger value of $d\sigma/d \mtt$. This larger cross section is due the fact that the NNLO result for $d\sigma/d \mtt$ at high \mtt is a decreasing function of the central scale. Indeed, the \ms prediction with a fixed \ms mass uses the central scale $\mu_0=\mbar$, while the prediction with the running mass uses a larger
value, ($\mu_0 \sim \mtt/2$) of the central scale. A qualitatively similar scale dependence of $d\sigma/d \mtt$ in the 
high-\mtt region (e.g., $\mtt \gtap 1$~TeV) is observed in the NNLO results~\cite{Czakon:2016dgf}
in the pole scheme.
Therefore, such scale dependence (both in the \ms and the pole scheme) is mostly driven by the scale \muR of $\as$ and the scale
\muF of the PDFs (increasing \mtt the PDFs are sensitive to the region of increasingly higher values of momentum fractions, where scaling violations are negative).

Our QCD predictions with running \ms mass use the central scale $\mu_0= \mu_k/2 \sim \mtt/2$ for all
the auxiliary scales \muR, \muF and \mum.
To disentangle the effect of the running of the top-quark mass $m_t(\mum)$ from the effect due to the running
of $\as$ and to the scaling violations of the PDFs, we also present 
(see the dashed lines in the lower panel of Fig.~\ref{fig:mtt-CMS-mt} right)
the results that are obtained by keeping $\mum=\mbar$ fixed, 
while still using the dynamic scale $\mu_k/2$ for \muR and \muF.
Even though the use of the running mass $m_t(\mu_k/2)$ leads to a slightly better agreement with the data,
the difference from the result with $\mum=\mbar$ 
(solid lines in Fig.~\ref{fig:mtt-CMS-mt} right)
is very small in comparison to the size of the theoretical and experimental uncertainties.

Comparing the QCD results at central scales in the lower panel of Fig.~\ref{fig:mtt-CMS-mt} right, we note the following features.
Going from $m_t(\mu_k/2)$ (solid lines) to $m_t(\mbar)$ (dashed lines), the \ms mass $m_t(\mum)$ of the top-quark increases
and, accordingly with naive expectations, the $\mtt$ cross section decreases. The decrease of the cross section is rather uniform
throughout, from the region of low to high values of $\mtt$. This is also not unexpected.
At low invariant masses the parametric dependence of the $\mtt$ cross section on $m_t(\mum)$ is large
(since the value of \mum effectively changes the position of the invariant-mass threshold), but the difference between the fixed and the dynamic scales, $\mum=\mbar$ and $\mum=\mu_k/2 \sim \mtt/2$, is very small. 
At high invariant masses this difference becomes larger, but
the parametric dependence of the $\mtt$ cross section on the mass $m_t(\mum)$ is much smaller.

In the highest invariant-mass bin of Fig.~\ref{fig:mtt-CMS-mt}, we also see that the difference between the solid and the dashed lines of
Fig.~\ref{fig:mtt-CMS-mt} (right) is definitely smaller than the difference between the central predictions (solid lines) on the left-hand side and
the right-hand side of Fig.~\ref{fig:mtt-CMS-mt}. This fact confirms our previous conclusion that the difference between our fixed-mass and running-mass predictions of Fig.~\ref{fig:mtt-CMS-mt} is mostly driven by the scales \muR and \muF of $\as$ and PDFs.
The quantitative dependence on the mass renormalisation scale is smaller for a twofold reason: 
the running of the \ms mass of the top quark is \textit{slower} than that of the QCD coupling $\as$
(see Eq.~(\ref{eq:slow})), and the scaling violations of the PDFs
increase with increasing $\mtt$.

From our discussion of the results in Fig.~\ref{fig:mtt-CMS-mt} we conclude that the data of Ref.~\cite{Sirunyan:2019jyn}
are not able to pin down effects produced by the running of the \ms mass in the NNLO predictions.
This conclusion is the consequence of the relatively large theoretical uncertainties of the predictions and, partly, of the size of the experimental errors. 

We complete our discussion on the invariant-mass distributions with few additional considerations.

The computation of QCD observables that depend on a single hard-scattering scale 
(e.g, the \ttb total cross section, which depends on the top-quark mass)
is usually performed by setting the central scale $\mu_0$ of the corresponding fixed-order calculation
to a value of the order of the hard-scattering scale.
The differential cross section $d\sigma/d \mtt$ is not a single-scale QCD observable, since it depends on \textit{both}
the top-quark mass and the invariant mass of the \ttb pair. If these two mass scales are parametrically different, we are dealing with a two-scale QCD observable, and high-order  radiative QCD corrections are expected to be quantitatively relevant.
In the context of fixed-order QCD predictions for two-scale observables, a customary procedure to investigate higher-order effects
is to set the auxiliary scale $\mu_0$ of central QCD predictions at a value within the range of the two mass scales.
In our predictions of  Fig.~\ref{fig:mtt-CMS-mt} with fixed and running \ms masses, we have considered the two `extreme'
choices $\mu_0=\mbar$ and $\mu_0 \simeq \mtt/2$, and we have found relatively similar (within scale uncertainties) results
for $\mtt \ltap 1$~TeV (i.e., the value of $\mu_k$ in the $4^{\text{th}}$ bin of Fig.~\ref{fig:mtt-CMS-mt}).

In the case of two-scale QCD observables, the use of fixed-order calculations with dynamic values of $\mu_0$ is expected to be a sensible theoretical procedure, provided the two scales are parametrically not \textit{very} different. The direct calculation of higher-order contributions 
(for instance, through all-order resummation techniques) is instead theoretically more appropriate 
in the kinematical region where the two mass scales are parametrically very different. In the specific case of
$d\sigma/d \mtt$, this is the multi-TeV invariant-mass region. In Ref.~\cite{Czakon:2018nun}, the differential cross section
$d\sigma/d \mtt$ at high (multi-TeV) values of $\mtt$ was studied by combining the NNLO calculation in the pole scheme
\cite{Czakon:2016dgf} with resummed calculations~\cite{Ahrens:2010zv,Ferroglia:2012ku} of contributions due to soft and collinear radiation. The combination of resummed calculations with the NNLO calculation in the \ms scheme can be of interest to perform
further investigations on the effects of the running of the top-quark mass.

\section{Summary}
\label{sec:summa}

This paper has been devoted to present and discuss QCD predictions for 
\ttb
production at the LHC by using the \ms renormalisation scheme for the definition of the top-quark mass.

We have remarked that the LHC experimental data refer to the production of `physical' (though unstable) top quarks and antiquarks
with a definite value of the pole mass
\Mt
(and width $\Gamma_t$).
At the theoretical level, this physical picture directly leads (in the limit $\Gamma_t \to 0$)
to considering perturbative calculations for on-shell \ttb production in the pole scheme.
We have then discussed how the on-shell calculations in the pole scheme can be transformed into corresponding
calculations in the \ms scheme through a formal all-order perturbative replacement of the renormalised top-quark mass.
We have highlighted possible unphysical features (e.g., in connection with  \ttb production thresholds)
that are produced by such formal replacement order-by-order in QCD perturbation theory.
We have also discussed how running-mass effects can be introduced in the perturbative calculation within the \ms scheme.

In the previous literature, QCD predictions within the \ms scheme had been limited to calculations of the \ttb total cross section up to NNLO and of single-differential cross sections up to NLO.
In this work we have computed
the total cross section and
single-differential distributions for \ttb production at NNLO 
by using the \ms scheme for the renormalisation of the top-quark mass.
The NNLO results substantially increase the precision of the 
\ms scheme theoretical results for differential distributions that were previously available only at NLO,
and, therefore, our results are
very relevant 
in the context of the experimental determination of the \ms mass of the top quark.

In our computation we have consistently included variations of the renormalisation scale \mum of the \ms mass
$m_t(\mum)$. We have then considered a 15-point variation 
(once variations of \muR and \muF are included)
of the QCD auxiliary scales (\muR, \muF and \mum) to estimate the scale uncertainties 
of the fixed-order QCD predictions. In particular, we find that 
the inclusion of \mum variations is crucial to consistently and correctly assess the size of the perturbative uncertainties 
of the \ms scheme predictions, especially at low perturbative orders and for differential distributions
(e.g., the distribution of the invariant mass, 
\mtt, 
of the \ttb pair) in the vicinity of \ttb production thresholds.
Moreover, we have also computed the invariant-mass distribution at high \mtt
by using a dynamic value of the renormalisation scale \mum, therefore effectively introducing a running value,
$m_t(\mum)$, of the top-quark mass in the \ms scheme.

We have presented the results of detailed QCD calculations of the \ttb total cross section 
in both the pole and \ms schemes up to NNLO by using different central values $\mu_0$
of the auxiliary scales \muR, \muF and \mum, and including uncertainties from scale variations
around the central scale. The comparison between the pole and \ms schemes shows consistent results within scale uncertainties.
Comparing the pole scheme results with $\mu_0=\Mt$ to the \ms scheme results with $\mu_0=\mbar$, 
we confirm previous findings in the literature: 
the perturbative convergence of these \ms scheme results  appears to be faster, 
with larger overlap of the scale uncertainty bands at subsequent perturbative orders,
and with smaller corrections and scale uncertainties at NNLO.
However, we have pointed out that the features of faster or slower apparent convergence strongly depend
on both the value of $\mu_0$ and the type of mass renormalisation scheme. Within each of the two 
mass renormalisation schemes, the apparent convergence can be made faster (or slower) by changing the value
of the central scale $\mu_0$.

Considering scales $\mu_0$ of the order of the top-quark mass,
we have presented results of single-differential distributions in the \ms scheme up to NNLO, 
and we have also compared them with corresponding results in the pole scheme. 
The comparison between the results in the two schemes leads to conclusions that are similar to those
that apply to the total cross section. In particular, we concluded that the shape differences between the pole and \ms scheme
results 
are significantly reduced by the inclusion of the high-order contributions, and they are quite small at NNLO.
Moreover, in all the kinematical regions that we have considered, we have noted a sizeable overlap between the pole and
\ms scheme uncertainty bands at NNLO. The high similarity between these pole and \ms scheme results
at NNLO for differential distributions is a relevant feature, since it also justifies the study of running-mass effects
through the introduction of dynamic values of the renormalisation scale \mum for the \ms mass $m_t(\mum)$.

We have considered a recent measurement of the \mtt differential cross section performed by the CMS Collaboration.
The measurement extends up to $\mtt \ltap 1$~TeV. We have computed corresponding QCD predictions
up to NNLO within the \ms scheme by using either fixed ($\mum=\mbar$) or dynamic ($\mum \sim \mtt/2$)
central values of \mum. We have discussed the effects that are produced by the dynamic scale.
We have observed  an excellent agreement between the experimental data and the theory results at NNLO.
In particular, the NNLO results lead to a sizeable reduction of the scale uncertainties with respect to the corresponding results
at NLO, thus paving the way to a precise determination of the top-quark mass in the \ms scheme.
The NNLO predictions with fixed and dynamic values of \mum are consistent within their scale uncertainties, whose size is similar 
to that of the experimental errors. Therefore, we have concluded that these CMS data are not able to pin down effects produced by the running of the \ms mass in the NNLO predictions. Additional theoretical studies of running-mass effects in QCD predictions are left to future
investigations.

\vskip 0.5cm
\noindent {\bf Acknowledgements}                                                                                                                                                         
                                                                                                                                                                                         
\noindent We wish to thank Paolo Nason for valuable comments and useful discussions.                                                                                              
This work is supported in part by the Swiss National Science Foundation (SNF) under contracts 200020$\_$169041 and 200020$\_$188464. The work of SK is supported by the ERC Starting Grant 714788 REINVENT.

\bibliography{biblio}

\providecommand{\href}[2]{#2}\begingroup\raggedright\begin{thebibliography}{10}

\bibitem{Azzi:2019yne}
P.~Azzi et~al., {\it {Report from Working Group 1}},  {\em CERN Yellow Rep.
  Monogr.} {\bf 7} (2019) 1--220, [\href{http://arxiv.org/abs/1902.04070}{{\tt
  arXiv:1902.04070}}].

\bibitem{Hoang:2020iah}
A.~H. Hoang, {\it {What is the Top Quark Mass?}},
  \href{http://arxiv.org/abs/2004.12915}{{\tt arXiv:2004.12915}}.

\bibitem{Baernreuther:2012ws}
P.~B{\"a}rnreuther, M.~Czakon, and A.~Mitov, {\it {Percent Level Precision
  Physics at the Tevatron: First Genuine NNLO QCD Corrections to $q \bar{q} \to
  t \bar{t} + X$}},  {\em Phys. Rev. Lett.} {\bf 109} (2012) 132001,
  [\href{http://arxiv.org/abs/1204.5201}{{\tt arXiv:1204.5201}}].

\bibitem{Czakon:2012zr}
M.~Czakon and A.~Mitov, {\it {NNLO corrections to top-pair production at hadron
  colliders: the all-fermionic scattering channels}},  {\em JHEP} {\bf 12}
  (2012) 054, [\href{http://arxiv.org/abs/1207.0236}{{\tt arXiv:1207.0236}}].

\bibitem{Czakon:2012pz}
M.~Czakon and A.~Mitov, {\it {NNLO corrections to top pair production at hadron
  colliders: the quark-gluon reaction}},  {\em JHEP} {\bf 01} (2013) 080,
  [\href{http://arxiv.org/abs/1210.6832}{{\tt arXiv:1210.6832}}].

\bibitem{Czakon:2013goa}
M.~Czakon, P.~Fiedler, and A.~Mitov, {\it {Total Top-Quark Pair-Production
  Cross Section at Hadron Colliders Through ${\cal O}(\alpha_s^4)$}},  {\em
  Phys. Rev. Lett.} {\bf 110} (2013) 252004,
  [\href{http://arxiv.org/abs/1303.6254}{{\tt arXiv:1303.6254}}].

\bibitem{Catani:2019iny}
S.~Catani, S.~Devoto, M.~Grazzini, S.~Kallweit, J.~Mazzitelli, and H.~Sargsyan,
  {\it {Top-quark pair hadroproduction at next-to-next-to-leading order in
  QCD}},  {\em Phys. Rev.} {\bf D99} (2019), no.~5 051501,
  [\href{http://arxiv.org/abs/1901.04005}{{\tt arXiv:1901.04005}}].

\bibitem{Czakon:2015owf}
M.~Czakon, D.~Heymes, and A.~Mitov, {\it {High-precision differential
  predictions for top-quark pairs at the LHC}},  {\em Phys. Rev. Lett.} {\bf
  116} (2016), no.~8 082003, [\href{http://arxiv.org/abs/1511.00549}{{\tt
  arXiv:1511.00549}}].

\bibitem{Czakon:2016ckf}
M.~Czakon, P.~Fiedler, D.~Heymes, and A.~Mitov, {\it {NNLO QCD predictions for
  fully-differential top-quark pair production at the Tevatron}},  {\em JHEP}
  {\bf 05} (2016) 034, [\href{http://arxiv.org/abs/1601.05375}{{\tt
  arXiv:1601.05375}}].

\bibitem{Czakon:2017dip}
M.~Czakon, D.~Heymes, and A.~Mitov, {\it {fastNLO tables for NNLO top-quark
  pair differential distributions}},
  \href{http://arxiv.org/abs/1704.08551}{{\tt arXiv:1704.08551}}.

\bibitem{Catani:2019hip}
S.~Catani, S.~Devoto, M.~Grazzini, S.~Kallweit, and J.~Mazzitelli, {\it
  {Top-quark pair production at the LHC: Fully differential QCD predictions at
  NNLO}},  {\em JHEP} {\bf 07} (2019) 100,
  [\href{http://arxiv.org/abs/1906.06535}{{\tt arXiv:1906.06535}}].

\bibitem{Langenfeld:2009wd}
U.~Langenfeld, S.~Moch, and P.~Uwer, {\it {Measuring the running top-quark
  mass}},  {\em Phys. Rev.} {\bf D80} (2009) 054009,
  [\href{http://arxiv.org/abs/0906.5273}{{\tt arXiv:0906.5273}}].

\bibitem{Ahrens:2011px}
V.~Ahrens, A.~Ferroglia, M.~Neubert, B.~D. Pecjak, and L.~L. Yang, {\it
  {Precision predictions for the t+t(bar) production cross section at hadron
  colliders}},  {\em Phys. Lett. B} {\bf 703} (2011) 135--141,
  [\href{http://arxiv.org/abs/1105.5824}{{\tt arXiv:1105.5824}}].

\bibitem{Dowling:2013baa}
M.~Dowling and S.-O. Moch, {\it {Differential distributions for top-quark
  hadro-production with a running mass}},  {\em Eur. Phys. J.} {\bf C74}
  (2014), no.~11 3167, [\href{http://arxiv.org/abs/1305.6422}{{\tt
  arXiv:1305.6422}}].

\bibitem{Sirunyan:2018goh}
{\bf CMS} Collaboration, A.~M. Sirunyan et~al., {\it {Measurement of the
  $t\overline{t}$ production cross section, the top quark mass, and the strong
  coupling constant using dilepton events in pp collisions at $\sqrt{s} =$ 13
  TeV}},  {\em Eur. Phys. J.} {\bf C79} (2019), no.~5 368,
  [\href{http://arxiv.org/abs/1812.10505}{{\tt arXiv:1812.10505}}].

\bibitem{Sirunyan:2019jyn}
{\bf CMS} Collaboration, A.~M. Sirunyan et~al., {\it {Running of the top quark
  mass from proton-proton collisions at $\sqrt{s} =$ 13 TeV}},  {\em Phys.
  Lett.} {\bf B803} (2020) 135263, [\href{http://arxiv.org/abs/1909.09193}{{\tt
  arXiv:1909.09193}}].

\bibitem{Catani:2007vq}
S.~Catani and M.~Grazzini, {\it {An NNLO subtraction formalism in hadron
  collisions and its application to Higgs boson production at the LHC}},  {\em
  Phys. Rev. Lett.} {\bf 98} (2007) 222002,
  [\href{http://arxiv.org/abs/hep-ph/0703012}{{\tt hep-ph/0703012}}].

\bibitem{Gray:1990yh}
N.~Gray, D.~J. Broadhurst, W.~Grafe, and K.~Schilcher, {\it {Three Loop
  Relation of Quark $\overline{\rm MS}$ and Pole Masses}},  {\em Z. Phys.} {\bf
  C48} (1990) 673--680.

\bibitem{Fleischer:1998dw}
J.~Fleischer, F.~Jegerlehner, O.~V. Tarasov, and O.~L. Veretin, {\it {Two loop
  QCD corrections of the massive fermion propagator}},  {\em Nucl. Phys.} {\bf
  B539} (1999) 671--690, [\href{http://arxiv.org/abs/hep-ph/9803493}{{\tt
  hep-ph/9803493}}]. [Erratum: Nucl. Phys.B571,511(2000)].

\bibitem{Chetyrkin:1999qi}
K.~G. Chetyrkin and M.~Steinhauser, {\it {The Relation between the
  $\overline{\rm MS}$ and the on-shell quark mass at order $\alpha_s^3$}},
  {\em Nucl. Phys.} {\bf B573} (2000) 617--651,
  [\href{http://arxiv.org/abs/hep-ph/9911434}{{\tt hep-ph/9911434}}].

\bibitem{Melnikov:2000qh}
K.~Melnikov and T.~v. Ritbergen, {\it {The Three loop relation between the
  $\overline{\rm MS}$ and the pole quark masses}},  {\em Phys. Lett.} {\bf
  B482} (2000) 99--108, [\href{http://arxiv.org/abs/hep-ph/9912391}{{\tt
  hep-ph/9912391}}].

\bibitem{Marquard:2016dcn}
P.~Marquard, A.~V. Smirnov, V.~A. Smirnov, M.~Steinhauser, and D.~Wellmann,
  {\it {$\overline{\rm MS}$-on-shell quark mass relation up to four loops in
  QCD and a general SU$(N)$ gauge group}},  {\em Phys. Rev.} {\bf D94} (2016),
  no.~7 074025, [\href{http://arxiv.org/abs/1606.06754}{{\tt
  arXiv:1606.06754}}].

\bibitem{Chetyrkin:1997dh}
K.~G. Chetyrkin, {\it {Quark mass anomalous dimension to ${\cal
  O}(\alpha_s^4)$}},  {\em Phys. Lett.} {\bf B404} (1997) 161--165,
  [\href{http://arxiv.org/abs/hep-ph/9703278}{{\tt hep-ph/9703278}}].

\bibitem{Vermaseren:1997fq}
J.~A.~M. Vermaseren, S.~A. Larin, and T.~van Ritbergen, {\it {The four loop
  quark mass anomalous dimension and the invariant quark mass}},  {\em Phys.
  Lett.} {\bf B405} (1997) 327--333,
  [\href{http://arxiv.org/abs/hep-ph/9703284}{{\tt hep-ph/9703284}}].

\bibitem{Baikov:2014qja}
P.~Baikov, K.~Chetyrkin, and J.~K{\"u}hn, {\it {Quark Mass and Field Anomalous
  Dimensions to ${\cal O}(\alpha_s^5)$}},  {\em JHEP} {\bf 10} (2014) 076,
  [\href{http://arxiv.org/abs/1402.6611}{{\tt arXiv:1402.6611}}].

\bibitem{Luthe:2016xec}
T.~Luthe, A.~Maier, P.~Marquard, and Y.~Schröder, {\it {Five-loop quark mass
  and field anomalous dimensions for a general gauge group}},  {\em JHEP} {\bf
  01} (2017) 081, [\href{http://arxiv.org/abs/1612.05512}{{\tt
  arXiv:1612.05512}}].

\bibitem{Baikov:2017ujl}
P.~Baikov, K.~Chetyrkin, and J.~K{\"u}hn, {\it {Five-loop fermion anomalous
  dimension for a general gauge group from four-loop massless propagators}},
  {\em JHEP} {\bf 04} (2017) 119, [\href{http://arxiv.org/abs/1702.01458}{{\tt
  arXiv:1702.01458}}].

\bibitem{Beneke:1994sw}
M.~Beneke and V.~M. Braun, {\it {Heavy quark effective theory beyond
  perturbation theory: Renormalons, the pole mass and the residual mass term}},
   {\em Nucl. Phys.} {\bf B426} (1994) 301--343,
  [\href{http://arxiv.org/abs/hep-ph/9402364}{{\tt hep-ph/9402364}}].

\bibitem{Bigi:1994em}
I.~I.~Y. Bigi, M.~A. Shifman, N.~G. Uraltsev, and A.~I. Vainshtein, {\it {The
  Pole mass of the heavy quark. Perturbation theory and beyond}},  {\em Phys.
  Rev.} {\bf D50} (1994) 2234--2246,
  [\href{http://arxiv.org/abs/hep-ph/9402360}{{\tt hep-ph/9402360}}].

\bibitem{Beneke:1994rs}
M.~Beneke, {\it {More on ambiguities in the pole mass}},  {\em Phys. Lett. B}
  {\bf 344} (1995) 341--347, [\href{http://arxiv.org/abs/hep-ph/9408380}{{\tt
  hep-ph/9408380}}].

\bibitem{Ayala:2014yxa}
C.~Ayala, G.~Cveti{\v c}, and A.~Pineda, {\it {The bottom quark mass from the $
  \boldsymbol{\Upsilon} (1S) $ system at NNNLO}},  {\em JHEP} {\bf 09} (2014)
  045, [\href{http://arxiv.org/abs/1407.2128}{{\tt arXiv:1407.2128}}].

\bibitem{Beneke:2016cbu}
M.~Beneke, P.~Marquard, P.~Nason, and M.~Steinhauser, {\it {On the ultimate
  uncertainty of the top quark pole mass}},  {\em Phys. Lett.} {\bf B775}
  (2017) 63--70, [\href{http://arxiv.org/abs/1605.03609}{{\tt
  arXiv:1605.03609}}].

\bibitem{Hoang:2017btd}
A.~H. Hoang, C.~Lepenik, and M.~Preisser, {\it {On the Light Massive Flavor
  Dependence of the Large Order Asymptotic Behavior and the Ambiguity of the
  Pole Mass}},  {\em JHEP} {\bf 09} (2017) 099,
  [\href{http://arxiv.org/abs/1706.08526}{{\tt arXiv:1706.08526}}].

\bibitem{FerrarioRavasio:2018ubr}
S.~Ferrario~Ravasio, P.~Nason, and C.~Oleari, {\it {All-orders behaviour and
  renormalons in top-mass observables}},  {\em JHEP} {\bf 01} (2019) 203,
  [\href{http://arxiv.org/abs/1810.10931}{{\tt arXiv:1810.10931}}].

\bibitem{Tanabashi:2018oca}
{\bf Particle Data Group} Collaboration, M.~Tanabashi et~al., {\it {Review of
  Particle Physics}},  {\em Phys. Rev.} {\bf D98} (2018), no.~3 030001.

\bibitem{Bonciani:2015sha}
R.~Bonciani, S.~Catani, M.~Grazzini, H.~Sargsyan, and A.~Torre, {\it {The $q_T$
  subtraction method for top quark production at hadron colliders}},  {\em Eur.
  Phys. J.} {\bf C75} (2015), no.~12 581,
  [\href{http://arxiv.org/abs/1508.03585}{{\tt arXiv:1508.03585}}].

\bibitem{Grazzini:2017mhc}
M.~Grazzini, S.~Kallweit, and M.~Wiesemann, {\it {Fully differential NNLO
  computations with MATRIX}},  {\em Eur. Phys. J.} {\bf C78} (2018), no.~7 537,
  [\href{http://arxiv.org/abs/1711.06631}{{\tt arXiv:1711.06631}}].

\bibitem{Catani:1996jh}
S.~Catani and M.~H. Seymour, {\it {The Dipole formalism for the calculation of
  QCD jet cross-sections at next-to-leading order}},  {\em Phys. Lett.} {\bf
  B378} (1996) 287--301, [\href{http://arxiv.org/abs/hep-ph/9602277}{{\tt
  hep-ph/9602277}}].

\bibitem{Catani:1996vz}
S.~Catani and M.~H. Seymour, {\it {A General algorithm for calculating jet
  cross-sections in NLO QCD}},  {\em Nucl. Phys.} {\bf B485} (1997) 291--419,
  [\href{http://arxiv.org/abs/hep-ph/9605323}{{\tt hep-ph/9605323}}]. [Erratum:
  Nucl. Phys. B510, 503 (1998)].

\bibitem{Catani:2002hc}
S.~Catani, S.~Dittmaier, M.~H. Seymour, and Z.~Trocsanyi, {\it {The Dipole
  formalism for next-to-leading order QCD calculations with massive partons}},
  {\em Nucl. Phys.} {\bf B627} (2002) 189--265,
  [\href{http://arxiv.org/abs/hep-ph/0201036}{{\tt hep-ph/0201036}}].

\bibitem{Cascioli:2011va}
F.~Cascioli, P.~Maierh{\"o}fer, and S.~Pozzorini, {\it {Scattering Amplitudes
  with Open Loops}},  {\em Phys. Rev. Lett.} {\bf 108} (2012) 111601,
  [\href{http://arxiv.org/abs/1111.5206}{{\tt arXiv:1111.5206}}].

\bibitem{Buccioni:2017yxi}
F.~Buccioni, S.~Pozzorini, and M.~Zoller, {\it {On-the-fly reduction of open
  loops}},  {\em Eur. Phys. J.} {\bf C78} (2018), no.~1 70,
  [\href{http://arxiv.org/abs/1710.11452}{{\tt arXiv:1710.11452}}].

\bibitem{Buccioni:2019sur}
F.~Buccioni, J.-N. Lang, J.~M. Lindert, P.~Maierh{\"o}fer, S.~Pozzorini,
  H.~Zhang, and M.~F. Zoller, {\it {OpenLoops 2}},  {\em Eur. Phys. J.} {\bf
  C79} (2019), no.~10 866, [\href{http://arxiv.org/abs/1907.13071}{{\tt
  arXiv:1907.13071}}].

\bibitem{Baernreuther:2013caa}
P.~B{\"a}rnreuther, M.~Czakon, and P.~Fiedler, {\it {Virtual amplitudes and
  threshold behaviour of hadronic top-quark pair-production cross sections}},
  {\em JHEP} {\bf 02} (2014) 078, [\href{http://arxiv.org/abs/1312.6279}{{\tt
  arXiv:1312.6279}}].

\bibitem{Ball:2017nwa}
{\bf NNPDF} Collaboration, R.~D. Ball et~al., {\it {Parton distributions from
  high-precision collider data}},  {\em Eur. Phys. J.} {\bf C77} (2017), no.~10
  663, [\href{http://arxiv.org/abs/1706.00428}{{\tt arXiv:1706.00428}}].

\bibitem{Schmidt:2012az}
B.~Schmidt and M.~Steinhauser, {\it {CRunDec: a C++ package for running and
  decoupling of the strong coupling and quark masses}},  {\em Comput. Phys.
  Commun.} {\bf 183} (2012) 1845--1848,
  [\href{http://arxiv.org/abs/1201.6149}{{\tt arXiv:1201.6149}}].

\bibitem{Aliev:2010zk}
M.~Aliev, H.~Lacker, U.~Langenfeld, S.~Moch, P.~Uwer, and M.~Wiedermann, {\it
  {HATHOR: HAdronic Top and Heavy quarks crOss section calculatoR}},  {\em
  Comput. Phys. Commun.} {\bf 182} (2011) 1034--1046,
  [\href{http://arxiv.org/abs/1007.1327}{{\tt arXiv:1007.1327}}].

\bibitem{Stevenson:1981vj}
P.~M. Stevenson, {\it {Optimized Perturbation Theory}},  {\em Phys. Rev.} {\bf
  D23} (1981) 2916.

\bibitem{Czakon:2016dgf}
M.~Czakon, D.~Heymes, and A.~Mitov, {\it {Dynamical scales for multi-TeV
  top-pair production at the LHC}},  {\em JHEP} {\bf 04} (2017) 071,
  [\href{http://arxiv.org/abs/1606.03350}{{\tt arXiv:1606.03350}}].

\bibitem{Hagiwara:2008df}
K.~Hagiwara, Y.~Sumino, and H.~Yokoya, {\it {Bound-state Effects on Top Quark
  Production at Hadron Colliders}},  {\em Phys. Lett. B} {\bf 666} (2008)
  71--76, [\href{http://arxiv.org/abs/0804.1014}{{\tt arXiv:0804.1014}}].

\bibitem{Kiyo:2008bv}
Y.~Kiyo, J.~H. Kuhn, S.~Moch, M.~Steinhauser, and P.~Uwer, {\it {Top-quark pair
  production near threshold at LHC}},  {\em Eur. Phys. J. C} {\bf 60} (2009)
  375--386, [\href{http://arxiv.org/abs/0812.0919}{{\tt arXiv:0812.0919}}].

\bibitem{Ju:2019mqc}
W.-L. Ju, G.~Wang, X.~Wang, X.~Xu, Y.~Xu, and L.~L. Yang, {\it {Invariant-mass
  distribution of top-quark pairs and top-quark mass determination}},
  \href{http://arxiv.org/abs/1908.02179}{{\tt arXiv:1908.02179}}.

\bibitem{Ju:2020otc}
W.-L. Ju, G.~Wang, X.~Wang, X.~Xu, Y.~Xu, and L.~L. Yang, {\it {Top quark pair
  production near threshold: single/double distributions and mass
  determination}},  \href{http://arxiv.org/abs/2004.03088}{{\tt
  arXiv:2004.03088}}.

\bibitem{Alekhin:2018pai}
S.~Alekhin, J.~Bl{\"u}mlein, and S.~Moch, {\it {NLO PDFs from the ABMP16 fit}},
   {\em Eur. Phys. J.} {\bf C78} (2018), no.~6 477,
  [\href{http://arxiv.org/abs/1803.07537}{{\tt arXiv:1803.07537}}].

\bibitem{Alekhin:2017kpj}
S.~Alekhin, J.~Bl{\"u}mlein, S.~Moch, and R.~Placakyte, {\it {Parton
  distribution functions, $\alpha_s$, and heavy-quark masses for LHC Run II}},
  {\em Phys. Rev.} {\bf D96} (2017), no.~1 014011,
  [\href{http://arxiv.org/abs/1701.05838}{{\tt arXiv:1701.05838}}].

\bibitem{Czakon:2018nun}
M.~Czakon, A.~Ferroglia, D.~Heymes, A.~Mitov, B.~D. Pecjak, D.~J. Scott,
  X.~Wang, and L.~L. Yang, {\it {Resummation for (boosted) top-quark pair
  production at NNLO+NNLL' in QCD}},  {\em JHEP} {\bf 05} (2018) 149,
  [\href{http://arxiv.org/abs/1803.07623}{{\tt arXiv:1803.07623}}].

\bibitem{Ahrens:2010zv}
V.~Ahrens, A.~Ferroglia, M.~Neubert, B.~D. Pecjak, and L.~L. Yang, {\it
  {Renormalization-Group Improved Predictions for Top-Quark Pair Production at
  Hadron Colliders}},  {\em JHEP} {\bf 09} (2010) 097,
  [\href{http://arxiv.org/abs/1003.5827}{{\tt arXiv:1003.5827}}].

\bibitem{Ferroglia:2012ku}
A.~Ferroglia, B.~D. Pecjak, and L.~L. Yang, {\it {Soft-gluon resummation for
  boosted top-quark production at hadron colliders}},  {\em Phys. Rev.} {\bf
  D86} (2012) 034010, [\href{http://arxiv.org/abs/1205.3662}{{\tt
  arXiv:1205.3662}}].

\end{thebibliography}\endgroup

\end{document}